\documentclass[sigchi]{acmart}
\renewcommand\footnotetextcopyrightpermission[1]{} 
\AtBeginDocument{%
  \providecommand\BibTeX{{%
    \normalfont B\kern-0.5em{\scshape i\kern-0.25em b}\kern-0.8em\TeX}}}

\acmConference[]{}{}{}
\acmBooktitle{ , , ,}

\usepackage{multirow}
\usepackage{multicol}
\usepackage{tabu}
\usepackage{longtable}
\usepackage{xcolor}
\usepackage{lipsum}
\usepackage{calc}
\usepackage{graphicx}
\usepackage{verbatim}
\usepackage[utf8]{inputenc}
\usepackage[english]{babel}
\usepackage{makecell,tabularx}

\usepackage{siunitx}
\usepackage{setspace}
\usepackage{adjustbox}
\usepackage{xcolor}

\begin{document}

\title{Cognitive Production Systems: A Mapping Study}

\author{Bastian Deutschmann}
\affiliation{
  \institution{Faculty of Informatics/Mathematics\\HTW University of Applied Sciences}
  \streetaddress{Friedrich List Platz 1}
  \city{Dresden}
  \country{Germany}
}

\author{Javad Ghofrani}
\affiliation{
  \institution{Faculty of Informatics/Mathematics\\HTW University of Applied Sciences}
  \streetaddress{Friedrich List Platz 1}
  \city{Dresden}
  \country{Germany}
}

\author{Dirk Reichelt}
\affiliation{
  \institution{Faculty of Informatics/Mathematics\\HTW University of Applied Sciences}
  \streetaddress{Friedrich List Platz 1}
  \city{Dresden}
  \country{Germany}
}

\renewcommand{\shortauthors}{Deutschmann et al. }

\begin{abstract}
Production plants today are becoming more and more complicated through more automation and networking. It is becoming more difficult for humans to participate, due to higher speed and decreasing reaction time of these plants. Tendencies to improve production systems with the help of cognitive systems can be identified. The goal is to save resources and time. This mapping study gives an insight into the domain, categorizes different approaches and estimates their progress. Furthermore, it shows achieved optimizations and persisting problems and barriers. These representations should make it easier in the future to address concrete problems in this research field. Human-Machine Interaction and Knowledge Gaining/Sharing represent the largest categories of the domain. Most often, a gain in efficiency and maximized effectiveness can be achieved as optimization. The most common problem is the missing or only difficult generalization of the presented concepts.
 \end{abstract}




\maketitle
\thispagestyle{empty}
\section{Introduction}\label{sec:introduction}

Over several decades the area of production systems had to face many different requirements and renew itself over and over. At the beginning this production as field and its related systems evolved from batch production over series- and varietal production to mass production where this development does not end~\cite{Duguay.1997}. The produced quantity of different products increased a lot due to higher demand. Modern production systems have to deal with even more problems and requirements. The expected reaction time on changes goes to zero and manufacturing demands change from "mass production to mass customization"\cite{Tran.2019}. Unexpected changes result both from outside the production system (changed orders) and inside the system (variances in the production itself)~\cite{Jost.2020}. A reaction on such changes demands an expert who is able to adjust the production system for the new situation. An Adjustment costs time where the system can't bring the expected outcome and thereby is really expensive. A perfect automated production system would neither need any make-ready time nor an expert. It would realize and understand changed circumstances and adapt to it in real time. Cognitive production systems as an approach to solve above mentioned problems was investigated by different authors. According to Tran et al.\cite{Tran.2019} a Cognitive Production System (CPS) possesses "cognitive capabilities such as perception, reasoning, learning, and cooperation". Cognitive robots have the "ability to handle unexpected situations" \cite{VolkerKruegerFrancescoRovidaBjarneGrossmannRonaldPetrickMathewCrosbyArnaudCharzoule.2019}. Pfeifer et al.~\cite{PfeiferSchmittStemmerRollofSchneiderDoro.2010} describe the concept of Cognitive Production Metrology "as an innovative solution to increase the manufacturing efficiency within flexible production lines. This is intended to contribute directly to reducing the complexity of pilot production series, for speeding up the production start time and assuring a maximum quality level for the process and product in dynamic environments.". Cognitive production solutions should not only be solutions themselves but also support humans in "human-robot collaborative manufacturing" \cite{S.M.MizanoorRahman.}. 
\\
The knowledge about using these solution approaches is only scattered treated until now~(\cite{Moghaddam.2018}\cite{Raj.2019}\cite{NumanM.Durakbasa.}). The efficiency of such systems and which problems they can solve in practice is not investigated enough. It has to be found out to which extent CPS improve an automated manufacturing with constantly changing requirements. The motivation here is, to find research gaps in this particular area. This paper should also suggest needed topics for further research.

The goal of this mapping study is to investigate the different aspects of the research area "cognitive production systems" and to identify research gaps. To achieve this goal, we constructed the following research questions(RQs): 
\begin{itemize}
\item RQ1: What types of concepts exist for CPS?
\item RQ2: Are these types only theory and concepts or are some of them used in practice?
\item RQ3: Which advantages do they (CPS) have in production and which problems of traditional automated production do they really solve?
\item RQ4: What are the most relevant problems and barriers for CPS-prototypes?
\end{itemize}   
The answer of these questions should simplify further work in this area. This paper provides also information about the advancement of CPS.
remainder of this paper is organized as following: 
In Section \ref{sec:related-work} existing studies along the topic of this paper are introduced. The search process with the search strings and the discrete selection of papers with inclusion and exclusion criteria is shown in Section \ref{sec:search-method}. Section \ref{sec:analysis} analysis answers the research questions previously posed with the help of the selected papers and their categorizations. All points which are not considered in detail in this work, but which nevertheless offer a value for the field of research, are briefly mentioned in Section\ref{sec:discussion}. Finally, Section \ref{sec:conclusion-future-work} provides a summary of the findings and shows future research possibilities which can clarify and deepen open points. 

\section{Related Work}\label{sec:related-work}

Lots of papers describe the future challenges and requirements of modern manufacturing. Osterrieder et al.~\cite{Osterrieder.2019} are using the term "smart factory" in their literature review.  They define the smart factory as a "future state of a fully connected manufacturing system, mainly operating without human force by generating, transferring, receiving and processing necessary data to conduct all required tasks for producing all kinds of goods". This work aims deeper into the area and does not restrict the manufacturing itself to work without human help. The human-robot collaboration is used in cognitive production systems to fulfil specific jobs. 
Contrary to the paper of Sharp et al.~\cite{Sharp.2018} this work is more generalised. Sharp et al. prove a growing interest in the use of machine learning in production. They mark for instance "Nearest-Neighbour" and "Support-Vector-Machine" as popular algorithms in production. Machine learning focuses on autonomous knowledge gain of computers. The reasons why a computer reacts in different situation with different actions are usually hided and afterwards not explainable. CPS also consider the past and the future of certain processes as important. Decisions in these systems are only made through clear reasons that a human person can understand. According to Goebel~\cite{Goebel.2018} "an effective explanation helps the explainer cross a cognitive valley[...]". That is why we want explainable-AI for Users of CPS in the factory. 
Missing tools that "support for managing the various aspects and complexities involved in the transformation towards a smart industry" are described by Breivold~\cite{Breivold.2017}. He marked the following six topics as the key drivers for developing production environments with cloud and IoT systems:  This work delivers methods and constructs to reach cognitive production. We tried to identify rudiments in which future tools could take place. The implementation of a CPS introduces new challenges that have to be solved. 
Hervé et al.~\cite{Panetto.2019} describe the importance of human-machine collaboration in socio-technical systems in their paper. For the operator 4.0, which is also often mentioned in other works~(\cite{Zolotova.2018}, \cite{Ruppert.2018}, \cite{DavidRomeroJohanStahreThorstenWuestOvidiuNoranPeterBernusAsaFastBerglundDominic.}), technologies must be used to this operator even better and more efficiently with intelligent systems. 
Other concepts such as decision-making and self-organization are also emerging as important cornerstones in future cyber-physical production systems. There are overlaps to this work. These and other challenges that were defined and explained by Panetto et. al are addressed by cognitive production systems with different concepts and methods. 
The goal of mass-personalization is used as one main motivation in the critical review of Moghaddam et al.~\cite{Moghaddam.2018}. In addition manufacturing should provide "smart and sustainable products and services, and enable real-time adaption to customer demand". They criticize the mass of new paradigms for smart manufacturing and try to filter out the characteristics of this type of production. CPS should not become a new basic paradigm, but rather combine existing concepts and techniques with some limitations to meet the requirements of modern manufacturing.

\section{Search Method}\label{sec:search-method}
\subsection{Search}
After designing the research questions we thought of different possibilities to accomplish an accurate search of papers. The set of papers was achieved with the following combination of search keywords:

\begin{center}
\begin{tabular}{ |m{2cm}|m{15em}| } 
\hline
Keyword & Synonyms \\ 
\hline
\multirow{3}{10em}{cognitive} & intelligent\\
& smart\\ 
& artificial \\ 
\hline
\multirow{5}{10em}{production} & system \\
& production system \\
& manufacturing \\ 
& factory \\ 
& manufacturing system \\ 
\hline
\multirow{2}{10em}{industry 4.0} & industrial internet of things \\
& IIoT \\
\hline
\multirow{2}{10em}{cyber physical} & industrial \\
& computer integrated \\
& computer aided \\
\hline
\end{tabular}
\end{center}

The keywords were defined in a thought process. Both the papers from (\cite{Moghaddam.2018}\cite{Panetto.2019}\cite{Zolotova.2018}\cite{Goebel.2018}\cite{Breivold.2017}\cite{Sharp.2018}\cite{Osterrieder.2019}\cite{Ruppert.2018}) and general considerations from the research on Cognitive Production Systems influenced the definition of the search keywords. We exclusively used electronic online databases with the above-mentioned keywords. We only focused on the most important databases in our opinion:
\begin{itemize}
\item ACM
\item IEEE
\item Scopus
\item Elsevier
\item Web of Science
\end{itemize}   
We know that there are other important search engines, but we wanted to limit ourselves to five in order not to break the time frame. In order to get a good search result it was necessary to adapt the search string to fit well to the used database. Table ~\ref{tab:search-strings} shows every composed search string which was used for the different databases. The database "Elsevier" was not used with a search string. In this database, the keywords of the other databases were combined in different fields to achieve the most accurate search.

\begin{table}[ht]
\centering
\begin{adjustbox}{width=1\columnwidth}
\small
\begin{tabular}{c|l}
  \hline
  Database & Search String \\
  \hline \hline
\multirow{5}{*}{{\tiny ACM}} & \multirow{5}{*}{\parbox{7cm}{\tiny (cognitive OR artificial) AND (cyber physical OR industrial OR ("computer aided") AND (system OR "production system" OR manufacturing OR factory OR "manufacturing system") AND (industry 4.0 OR "industrial internet of things" OR IIoT OR CPPS OR digital OR automation) AND (collaborative OR adaptability OR "human-machine" OR collaboration OR "human machine collaboration")}} \\ \\ \\ \\ \\
\hline
\multirow{5}{*}{{\tiny IEEE}} & \multirow{5}{*}{\parbox{7cm}{\tiny (intelligent OR smart OR cognitive OR artificial) AND (cyber physical OR industrial OR "computer integrated" OR "computer aided") AND (system OR "production system" OR manufacturing OR factory OR "manufacturing system") AND (industry 4.0 OR "industrial internet of things" OR IIoT OR CPPS OR digital OR automation) AND (collaborative OR adaptability OR reconfiguration OR "human-machine" OR collaboration OR "human machine collaboration")}} \\ \\ \\ \\ \\
\hline
\multirow{5}{*}{{\tiny Scopus}} & \multirow{5}{*}{\parbox{7cm}{\tiny (ALL((intelligent  OR  smart  OR  cognitive  OR  artificial) AND (cyber AND physical OR industrial OR "computer integrated" OR "computer aided") AND (system OR "production system" OR manufacturing OR factory OR "manufacturing system") AND (industry 4.0 OR "industrial internet of things" OR IIoT OR CPPS OR digital OR automation) AND (collaborative OR adaptability OR reconfiguration OR "human-machine" OR collaboration OR "human machine collaboration")))}} \\ \\ \\ \\ \\
\hline
\multirow{5}{*}{{\tiny Web of Science}} & \multirow{5}{*}{\parbox{7cm}{\tiny ALL=((intelligent OR smart OR cognitive OR artificial AND (cyber physical OR industrial OR "computer integrated" OR "computer aided") AND (system OR "production system" OR manufacturing OR factory OR "manufacturing system") AND (industry 4.0 OR "industrial internet of things" OR IIoT OR CPPS OR digital OR automation) AND (collaborative OR adaptability OR reconfiguration OR "human-machine" OR collaboration OR "human machine collaboration"))}} \\ \\ \\ \\ \\
\hline
\end{tabular}
\end{adjustbox}
\caption{Used Search Strings in the Scientific Databases} 
\label{tab:search-strings}
\end{table} 

With the different search strings mentioned above 1255 papers could be found in the five search engines. A detailed analysis of the material found is provided in Section \ref{sec:analysis}.

\subsection{Selection Strategy}\label{subsec:selection-strategy}

The amount of papers that were found necessitated a selection strategy. In the following sections the inclusion and exclusion criteria is explained. The goal was to refine the set of papers to only the most relevant ones.

\subsubsection{Inclusion Criteria}\label{subsubsec:inclusion-criteria}
The inclusion criteria describes all features, that will lead to an inclusion of the paper in this study. First of all a paper has to have at least one logic combination of keywords in the title or the abstract. In addition the document type is an important criteria for us. Only contribution papers, conference papers, journal papers and workshop papers are included in this study. We assume that all papers from selected scientific datasets are already peer reviewed.
\subsubsection{Exclusion Criteria}\label{subsubsec:exclusion-criteria}
The following criteria defines when a paper is excluded from the study. All papers had to be released in or after 2015 to ensure up-to-dateness in the topic. It should also cover a period of five years in order to show trends over time. We only considered papers in English language. If the state of the work is editorial or conceptional it is also excluded. Short papers, contributions from students e.g. bachelor, master and PhD works, blogs, magazines are not part of this study. Studies that did not lead to an result or an empirical finding were also excluded.

\begin{figure}[h]
\caption{Paper by search engines}
 \label{fig:papers_search_enginnes}
\includegraphics[trim=2cm 10.1cm 2cm 11.5cm ,clip,width=0.45\textwidth]{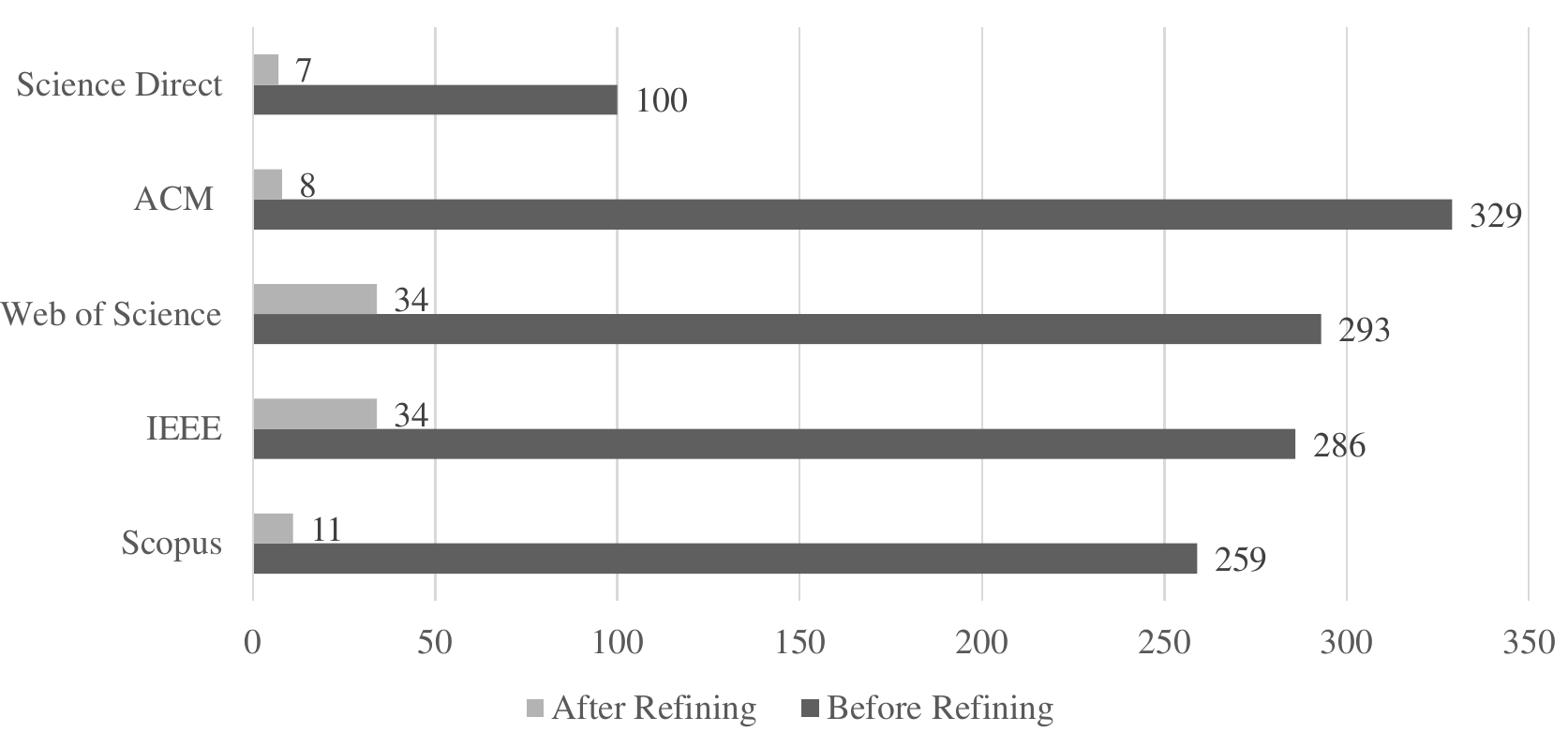}
\centering
\end{figure}

With the help of refining, the quantity of 1255 papers was reduced to 88. A full list of paper can be found on figshare. (DOI: 10.6084/m9.figshare.11907042) This corresponds to a share of 7.01\% of the total quantity. The distribution of the papers among the individual search engines before after refining can be seen in Figure \ref{fig:papers_search_enginnes}. It should be noted that 94 paper are shown after refining. The higher number results from six papers that were found in two different search engines and remained after the refining process.

\section{Analysis}\label{sec:analysis}
\begin{figure}[h]
\caption{Paper by years}
 \label{fig:papers_years}
\includegraphics[trim=2cm 10cm 2cm 11.2cm ,clip,width=0.45\textwidth]{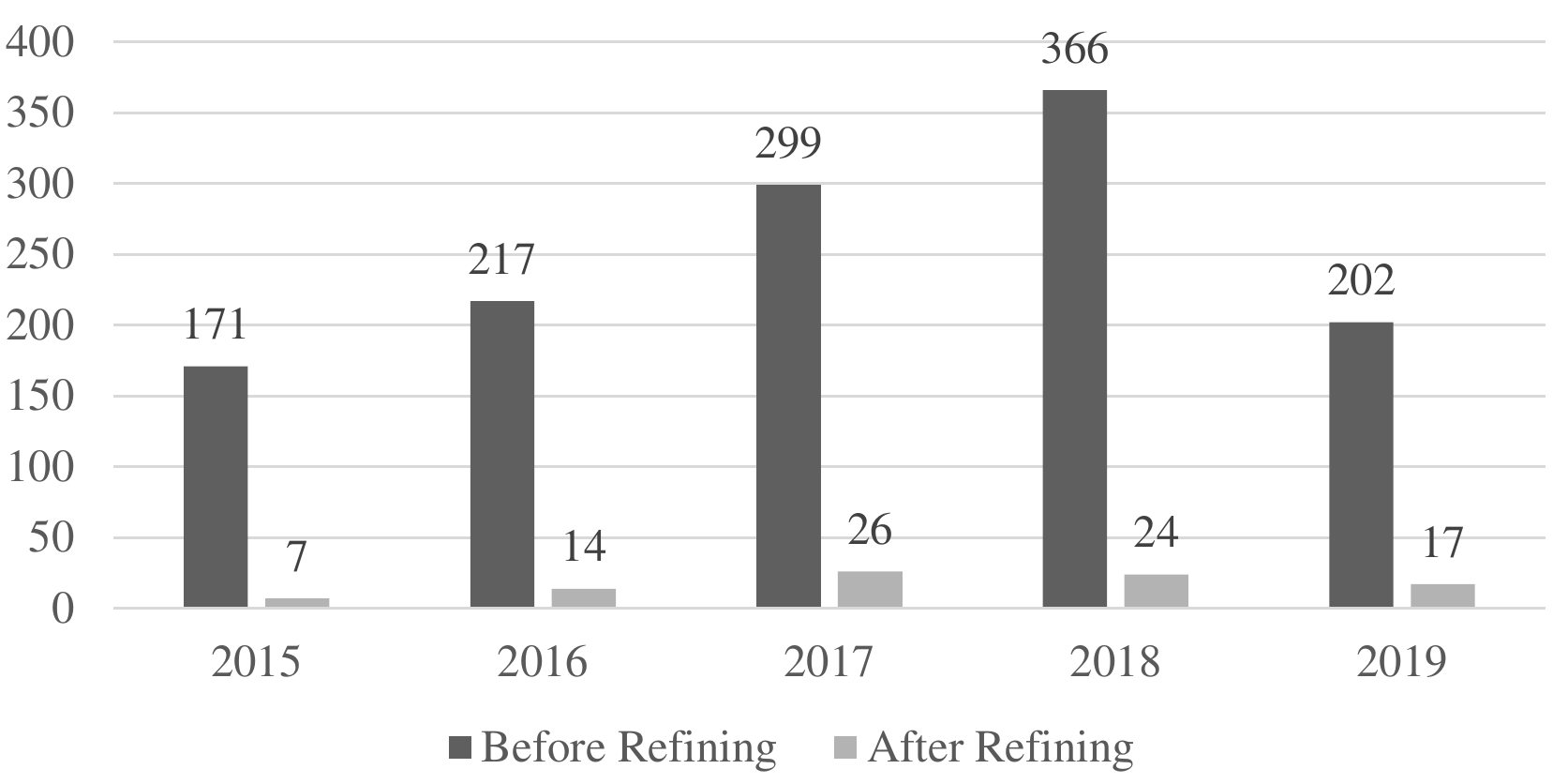}
\centering
\end{figure}

First of all we sorted the found works by years to check the relevance of the topic. The results of this sort divided in before and after refining are shown in figure \ref{fig:papers_years}.

The amount of paper found prior to refining increased sharply from 2015 onwards, suggesting a hype surrounding the subject. The small amount of work in 2019 can be explained by the timing of the mapping study. The search for the work was already completed in October 2019, which is why the quantity would be significantly higher if the study were to be repeated in 2020. Looking at the number of papers after refining, the continuing hype until 2019 is even clearer. Although only 202 papers were found by 2019, 17 survived the refining process. This means that the quotient between work before and after refining is highest this year.

The next step was to evaluate the sources from which the present work originated. In total 55 different sources were found. Of these, 39 sources with exactly one paper are part of this study. All sources with at least two papers contributed are shown in Figure 3.

\begin{figure*}[h]
\caption{Paper by sources}
 \label{fig:papers_sources}
\includegraphics[trim=2cm 10.5cm 3.2cm 11.8cm ,clip,width=0.98\textwidth]{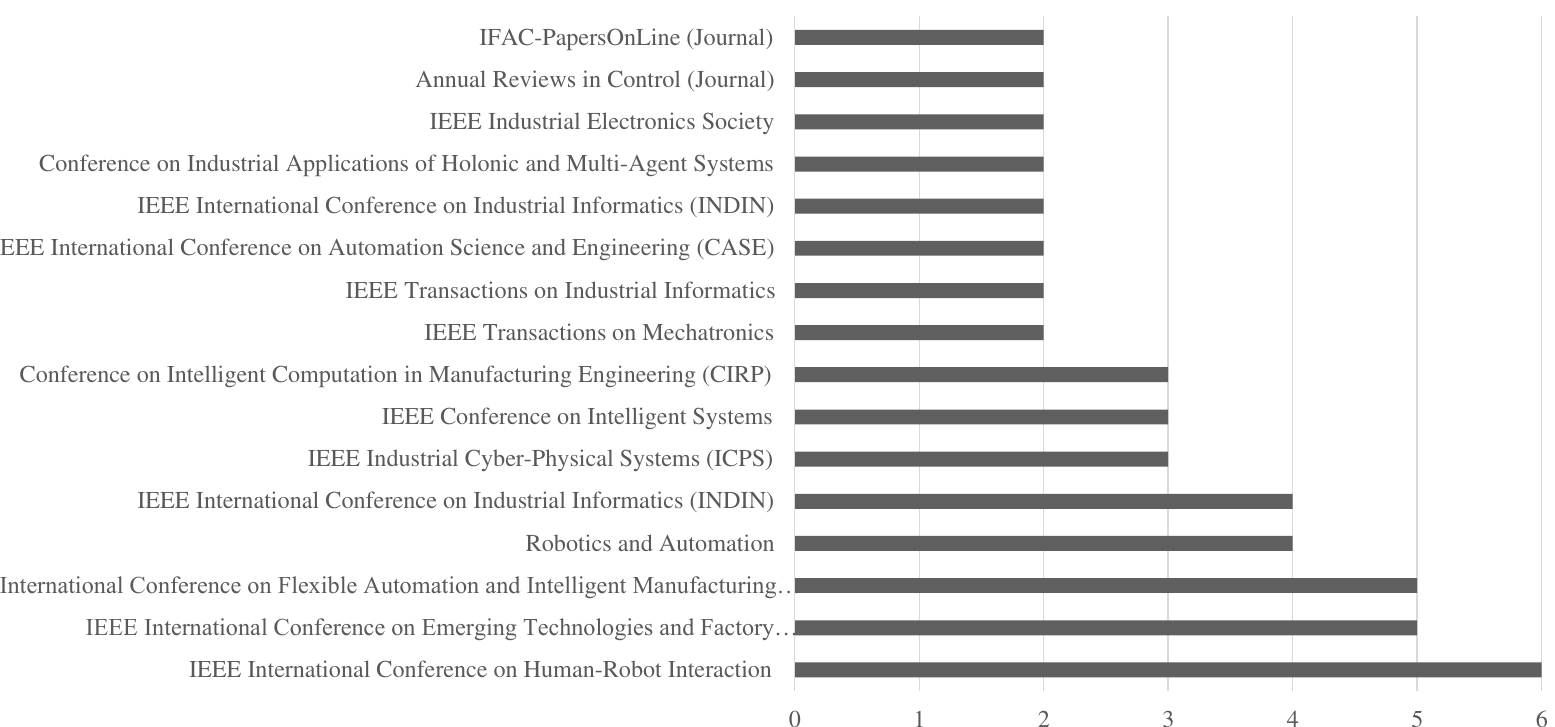}
\centering
\end{figure*}

\begin{figure*}[h]
\caption{Paper by author}
 \label{fig:papers_authors}
\includegraphics[trim=2cm 11.2cm 2cm 13cm ,clip,width=0.95\textwidth]{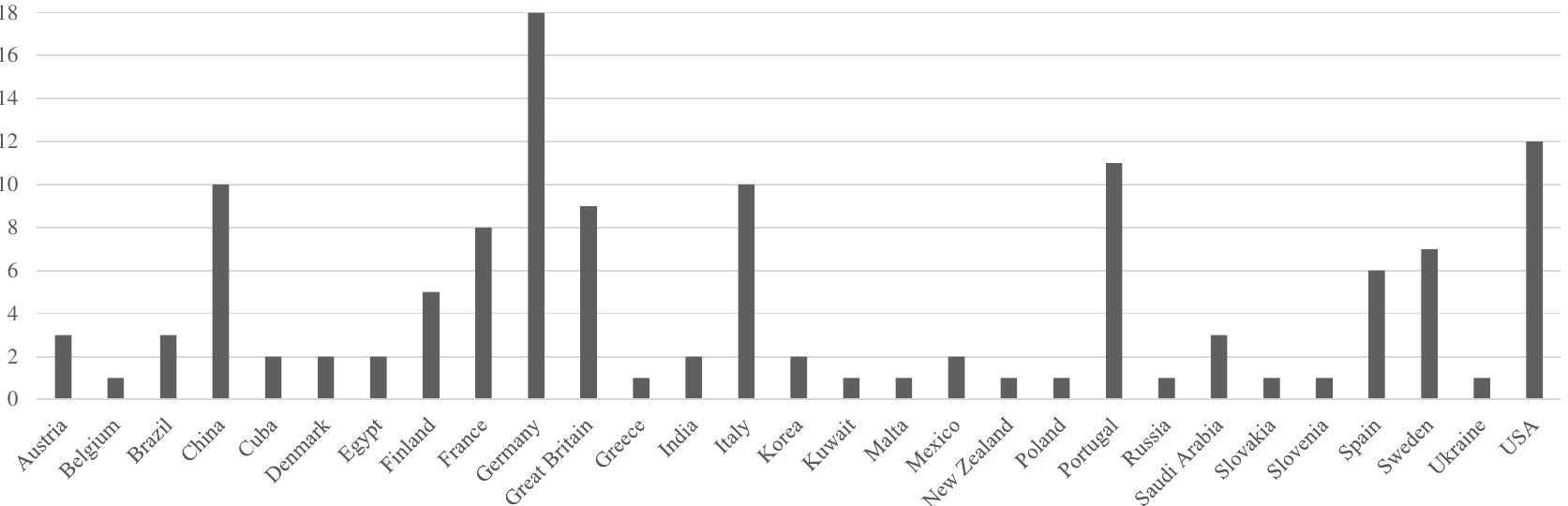}
\centering
\end{figure*}

The source with the most contributed papers already gives an outlook on an important finding that will later play a role in answering the research questions. The "IEEE International Conference of Human-Robot Interaction" provides a total of seven papers. The other major sources mainly deal with the topics of manufacturing automation, robots and intelligent systems. These subject areas match the selected search key words and thus confirm to some extent their relevance in research. If you look at the countries of the authors of the respective papers, you will find that some countries predominate. 
Figure \ref{fig:papers_authors} shows the number of authors per country. It should be noted that co-authors are also included in the count, which is why the sum of all bars is greater than 88. Furthermore, only the country listed in the paper is considered here. So it does not have to be the country of origin of the respective author or co-author, but can also only be the country of residence in which the research was conducted. 

With 18 mentions, Germany is the most frequent country followed by the USA, Portugal, Italy and China. While on the one hand countries like Germany were expected, mainly because of the topic "Industry 4.0", and on the other hand the USA and China, countries like Italy and Portugal surprised us. We have no concrete explanation for the frequency of the unexpected countries.

\begin{figure}[h]
\caption{RQ1: What types of of concepts exist for cognitive production systems?}
 \label{fig:1-ff}
\includegraphics[trim=2cm 9.6cm 2cm 11cm ,clip,width=0.45\textwidth]{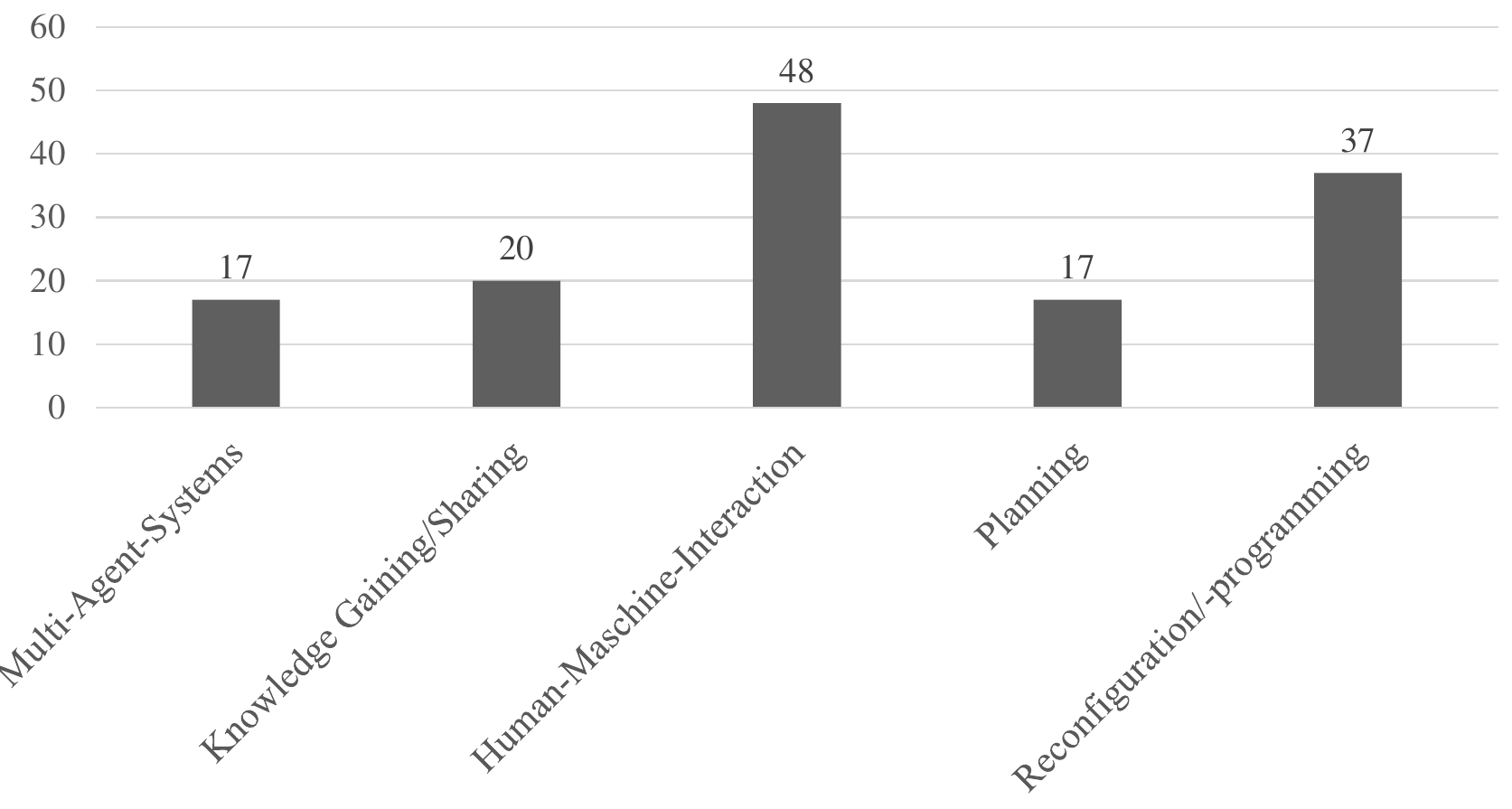}
\centering
\end{figure}

\paragraph{RQ1: What types of of concepts exist for cognitive production systems?}
After a general evaluation of the data stock, the concrete examination of the previously established research questions and all the work that has survived the refining process follows.
To answer the first research question, we have divided all papers into concept categories. Concepts are abstract solutions which can then be implemented in reality by means of methods.
The different concept categories were not defined in advance. They arose directly from the analysis process of the individual papers. If a type appeared several times specifically as a mention or indirectly as a specialized subcategory, it was included in the categorization. In the process, individual types were discarded if they were too specific and did not categorize at least ten papers. In the end, five types were left, which could categorize all 88 papers. Figure \ref{fig:1-ff} shows how many papers were placed in each category.

Even the definition of the individual categories quickly reveals that individual terms are mutually dependent and are not free of overlaps. It was not our goal to assign each paper exactly to one type, but to find enough topics that cover all found works in the context of "Cognitive Production Systems". Due to this fact, single papers can cover several categories. In total, the 88 papers were categorized 139 times. The largest topic is clearly "Human-Machine Interaction" with 48 categorized papers. Here, the already mentioned source "IEEE International Conference of Human-Robot Interaction" from which most of the papers originate, becomes clear again. This category includes all papers that in any way explore the interaction between humans and machines in the context of Cognitive Production Systems. 
Topics such as mechanical cooperation \cite{AlessandroRonconeOlivierManginBrianScassellati.2017}, communication or information exchange between humans and machines \cite{Emmanouilidis.2019} or gadgets for humans, which are supposed to facilitate cooperation, are described here\cite{Murauer.2019}. A total of 37 papers were assigned to the category "reconfiguration/programming". This category deals with changing production processes to which reprogramming and reconfiguration should be used to react dynamically. 
Topics are for example the creation of reconfigurable plans generated from a specific demand \cite{Wan.2019}, the reconfiguration of individual robotic agents \cite{Soumya.2018} or the description of a versatile gripper as basic requirement for reconfigurable production systems \cite{Borisov.2018}. 
The three other categories are represented with similar frequency. They are "Planning", "Multi-Agent-Systems" and "Knowledge Gaining/Sharing". The type "Planning" includes papers that deal with the planning of production processes. It does not matter when planning takes place or how long it takes. There are mainly approaches that plan in real time during the running production, incorporate newly created data into the planning and revise the plan \cite{Sadik.2017}\cite{AmedeoCestaAndreaOrlandiniGiulioBernardiAlessandroUmbrico.2016}\cite{AlessandroRonconeOlivierManginBrianScassellati.2017}.

"Multi-Agent-Systems" describes production environments in which a large number of different agents work on a superordinate goal and achieve this goal through division of labour and cooperation. In many cases, in addition to a large number of machines, a certain number of human participants interact with each other. Among other things, "cloud manufacturing" is used in this research field to better distribute the various production tasks between the individual agents \cite{Sarkar.2018}. 
In addition, there are approaches to visualize the communication between individual agents and to simplify it for the human employee by means of a monitoring system \cite{Mayer.2017}\cite{J.Jokinen.2016}. Different areas of artificial intelligence such as machine learning are also applied. By means of machine learning algorithms, decision support is offered for the coordination of a multitude of machines~\cite{Wang.2016}. The last category is called "Knowledge Gaining/Sharing". Papers in this category deal with the extraction of information from the running production processes. The origin of information can be very different. Not only machine data is gained, but also interpersonal communication is stored and metadata is generated (\cite{Bolmsjo.2015}\cite{HyeongGonJo.2017}\cite{Emmanouilidis.2019}). 
All of these information must also be sent to the right place at the right time and thus to the right machine or the right person. To simplify this problem some papers deal with the construction of a central information platform, which can centralize the extraction and exchange of information and simplify the distribution considerably~\cite{Alm.2015}. On the basis of this information, well-founded decisions can be made more easily and quickly by humans as well as automatically by machines \cite{Mabkhot.2016}.  

\paragraph{RQ:2 Are these types only theory and concepts or are some of them used in practice?}
In order to answer "Research Question 2", the categories of the papers classified in "Research Question 1" were evaluated in terms of their results. We decided on three possible assessments. The category "Only Concepts"(Category 1) includes papers that present a theoretical concept and describe how it works. If papers also show a selection of results from experiments carried out with the applied concept in reality, they are classified in the category "Concepts with Experiments"(Category 2). The category "Concepts with Case Study"(Category 3) is for those papers that apply their concept to a real production environment. Real test environments that do not directly participate in the value creation are also allowed. This categorization is intended to show the progress of the individual papers within the categories. Papers in category 1 are usually at the beginning of their research activity and later provide new results which we would classify in category 2 or 3 based on our model. Papers in category 3 have usually already carried out a real simulation and have thus tested their concepts and proven that they have an effect. In the following figure~\ref{fig:2-ff}, the five concept categories are evaluated according to their state of research.

\begin{figure}[h]
\caption{RQ:2 Are these types only theory and concepts or are some of them used in practice?}
 \label{fig:2-ff}
\includegraphics[trim=2cm 9.2cm 2cm 10.4cm ,clip,width=0.45\textwidth]{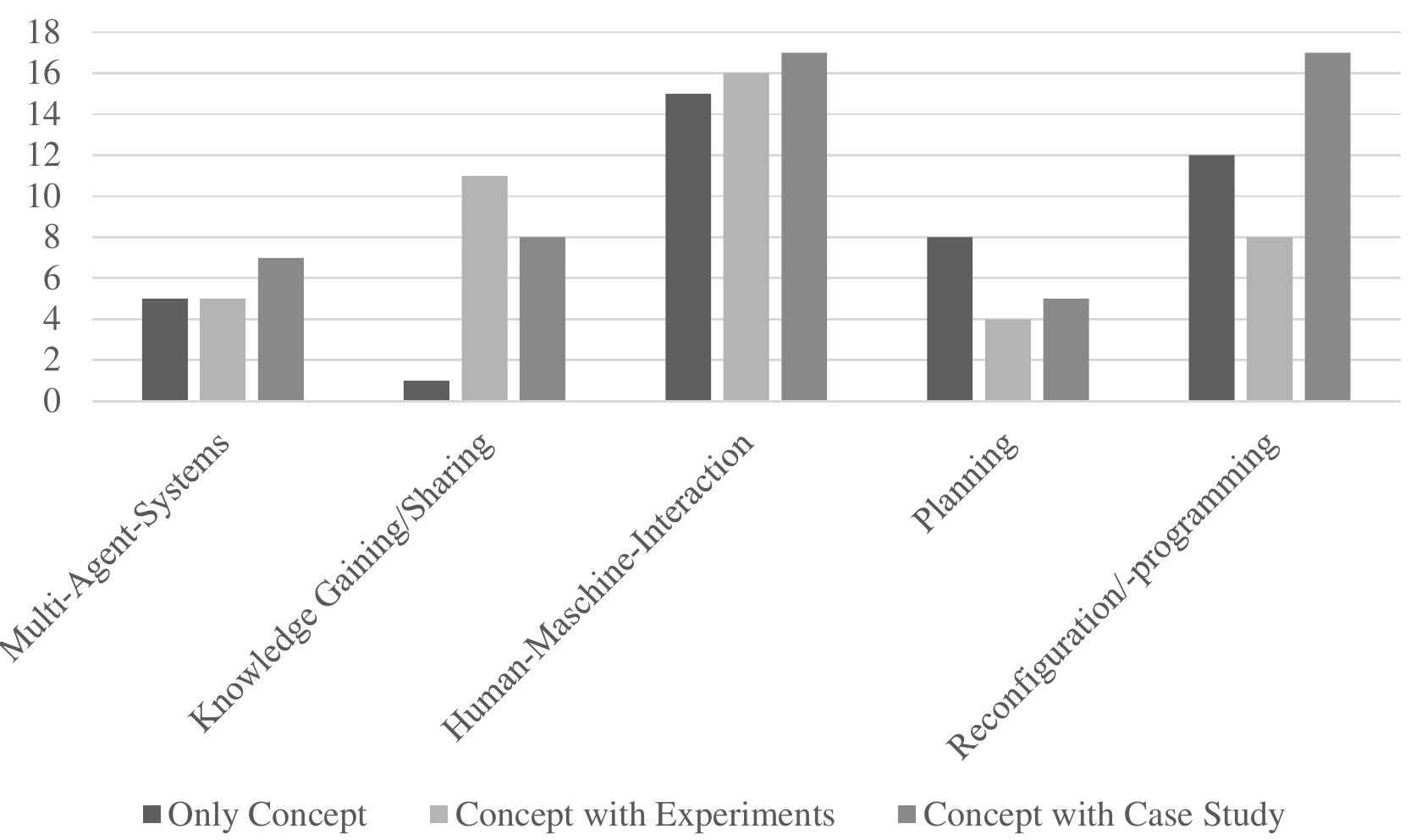}
\centering
\end{figure}

Among other things, the category "Knowledge-Gaining/-Sharing" is conspicuous here, which seems to be largely verified with real data. The category "Planning" seems to produce new theories in this period, which are not yet based on real data. The other categories are at different stages of development, but have a selection of papers at each stage. Similar to the distribution of papers by year, this indicates a continuing hype and interest in the topic. New theories and concepts are being developed, and existing concepts are being backed up with real data. 

\paragraph{RQ3: Which advantages do they (CPS) have in production and which problems of traditional automated production do they really solve?}
The answer to research question 3 compares the results of the individual papers. Here the focus is on the optimizations and advantages achieved by introducing new concepts. It should be noted that papers that are still in stage 1 "Only Concept" according to research question 2, do not provide any evidence to substantiate any advantages they may suggest. Nevertheless, these have been included in the statistics. In addition, some papers did not indicate any optimization or advantages at all and were therefore not included in the statistics. 
The following figure \ref{fig:3-ff} shows the total frequency of the achieved benefits. Since individual papers can also achieve more than one improvement with their concept, the sum of the advantages is greater than the sum of the papers. To count as a mention, an advantage may be mentioned directly or indirectly, but must be recognisable without a basic understanding of the concept.

\begin{figure}[h]
\caption{RQ3: Which advantages do they (CPS) have in production and which problems of traditional automated production do they really solve?}
 \label{fig:3-ff}
\includegraphics[trim=2cm 4.25cm 2.8cm 4.25cm ,clip,width=0.45\textwidth]{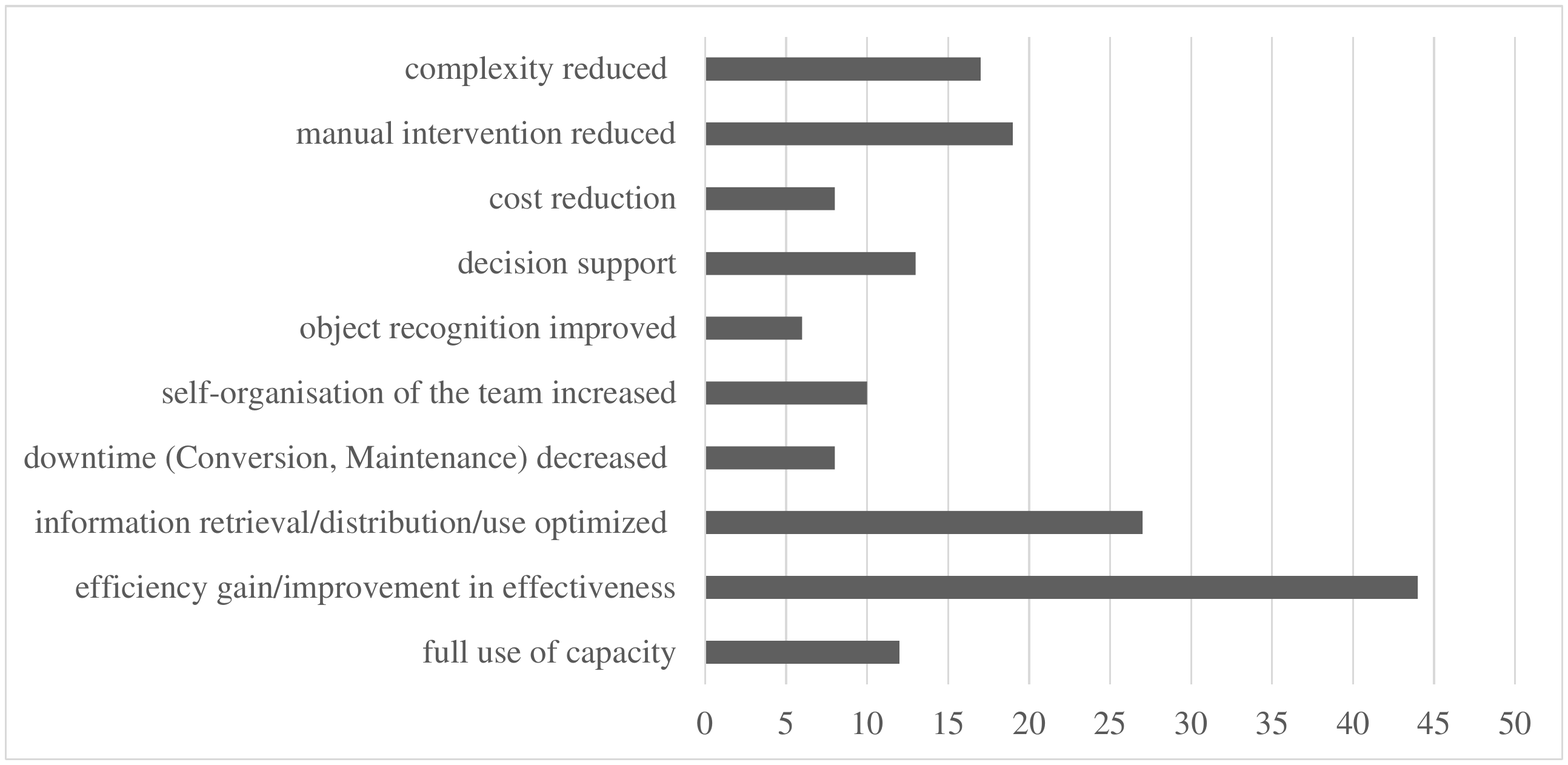}
\centering
\end{figure}

Figure \ref{fig:3-ff} shows a clear trend. A gain in efficiency and improvements in effectiveness is, with 44 mentions, the most frequently achieved optimisation through the application of the concepts in the context of "cognitive production systems". Most of the papers thus did justice to the objective addressed in the motivation. A main goal of CPS is to ensure even faster reactions to new product requests. In general, this is most easily achieved with maximized effectiveness. If, at the same time, costs can be saved through increased efficiency, such concepts can also increase the profit of production systems. This optimization is also evident in the point "cost reduction" with 13 mentions. The clear difference between the numbers of the two points shows that costs only played a subordinate role in most research, as the main objective was rather in other areas. However, there were no concrete details or key figures on cost savings. Rather, these result from other factors, such as savings in production time and reduction of downtime. For example, Dey and Sarkar~\cite{Dey.2016} address "maintenance [...][as] a crucial aspect to prevent untimely breakdown as it might lead to decrease in productivity and huge financial losses". 
Bewley and Liarokapis~\cite{TomBewley.2019} mention gamification as another way to reduce costs. Another important optimization with 27 mentions is the handling of information. Here, various methods are used to ensure better information management over the entire life cycle. The focus is on information from its generation and extraction, distribution and storage to its use and archiving. This shows once again how interlinked concept categories and obtained optimizations can be. For example, according to Emmanouilidis et al.~\cite{Emmanouilidis.2019} "more efficient integration of human and non-human actors in sociotechnical systems". According to Meyer et al.~\cite{OlgaMeyer.2018} the demand for standardization and open standards in the communication of information supports a better information management. The visualization of information by tools for the human employee can give him an overview of the information flood of production systems \cite{Mayer.2017}. 
Newly developed information models like the one developed by Mabkhot et al.\cite{Mabkhot.2016} can also enable and simplify other topics of cognitive production systems such as reconfiguration. Thus, improved information management is in many cases a basic requirement for the application of new concepts and methods and also influences other optimizations obtained from Figure~\ref{fig:3-ff} such as "manual intervention reduced" (19 mentions), "decision support" (13 mentions) and "self-organisation of the team increased" (10 mentions). Optimizations could also be developed in more specific areas. For example, an improved object recognition was mentioned 6 times, which helps machines to react to unprepared situations. This also includes the recognition of gestures of human operators, which is important for the cooperation between man and machine~\cite{Coupete.2016}. 
A positive side effect of some optimization is the improved integration of humans into the production process. For example, concepts can reduce manual intervention in production but at the same time better integrate humans by placing them in a supervisor level through tool and algorithmic support~\cite{PacauxLemoine.15.05.201818.05.2018}.   

\paragraph{RQ4: What are problems and barriers for CPS-prototypes?}
The fourth and final research question completes the frame around this mapping study. Here all barriers and problems of the papers were identified and categorized. It is important to note that this research question, like research question 3, depends on research question 2. In most cases real problems only become apparent during the first experiments or tests and during the real implementation of a concept. Since papers in state 1 (only concept) lack these results, most of them did not specify any concrete problems and limitations. The following figure~\ref{fig:4-ff} shows the problems and limitations of the 88 refined papers, divided into seven categories. The occurrence of a problem was treated like the development of an optimization from research question 3. All directly and indirectly mentioned facts were recorded and categorized.

\begin{figure}[h]
\caption{RQ4: What are problems and barriers for CPS-prototypes?}
 \label{fig:4-ff}
\includegraphics[trim=2cm 10.8cm 2.6cm 12cm ,clip,width=0.45\textwidth]{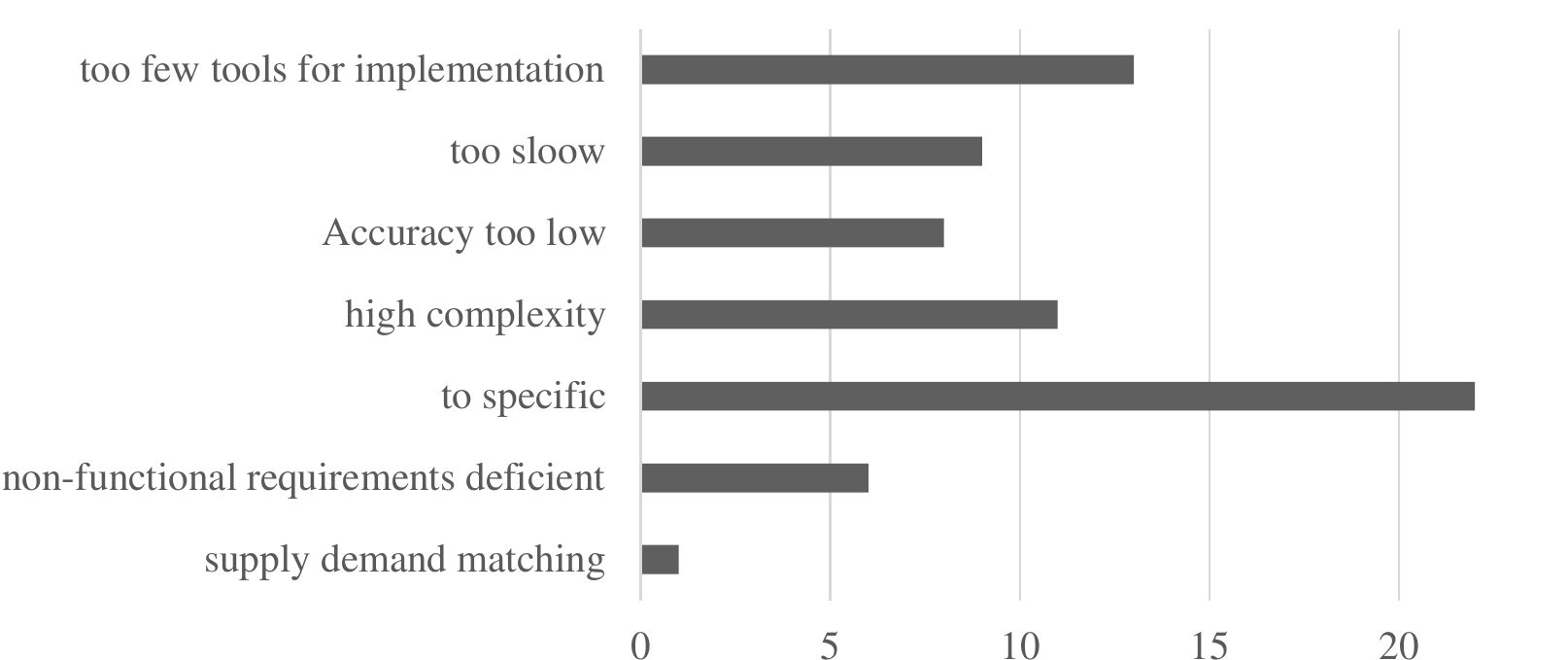}
\centering
\end{figure}
With 22 mentions, "too specific" is the most common problem. The problem "too specific" means above all the generalization of the concepts which is not or difficultly possible. Many of these concepts only work in their intended test environment and can only be applied to other use cases with difficulty or at great expense. This also means that solutions may not be flexible enough to react to small changes in the environment. In most cases, researchers want to use parts of their future work to generalize the specific results~(\cite{Peruzzini.2018}\cite{Askarpour.2017}\cite{Lee.2017}). If researchers still have to develop specific tools or applications that make the concepts successfully implementable in real industrial scenarios, they belong to the category "too few tools for implementation". 
Some papers call for new tools they need to continue working and others want to develop these applications themselves. Most of them are papers in state 1 and 2, because by definition the concept does not have a real implementation in an industrial use case. Three other common problems that also occur with other new developments are too low speed (computing time, production time, transfer time), too high inaccuracy (production, recognition, data collection, configuration) and too high complexity for the human operator. If the technology used is too complex, this drastically extends the training period and can also lead to comprehension problems for employees during operation, who then work more inefficiently. 
Problems with non-functional requirements such as data protection, security and user-friendliness occurred in six operations. Overall, it should be noted that most of the papers devote a very small amount of time to the problems or open issues. This is of course understandable, as in most cases a new concept with new findings is presented, which is the focus of attention. The problems identified and categorized were largely to be expected and fit the usual problems of new technical developments.

\section{Discussion}\label{sec:discussion}
 Despite the relatively large amount of papers found, some topics had to be excluded to increase the relevance of the results. The following topics survived the exclusion and inclusion criteria but were removed afterwards. Uber-Manufacturing refers to producing in a set of production facilities at different locations and distributing production requests as needed. We excluded this topic because we wanted to focus on the progress within a production facility. We decided to exclude paper dealing with Digital Twins, as most of them are only simulations. 
 We believe that digital twins alone have nothing to do with cognitive production systems. It is rather a phase and a tool to plan and produce the new systems. Nevertheless it has to be mentioned that Patented Intelligence has not been excluded. In theory it is a digital twin of a human decision making process. Since this Digital Twin is used as a decision model directly in production, this topic had sufficient relevance for us and belongs to the category Knowledge Gaining/Sharing and Human-Machine-Interaction. The term "learning factory" occurs relatively frequently. We distinguish ourselves from it with one aspect. Learning factories are primarily intended to teach people something and promote acceptance of new digital developments. We know that the training of employees and the sharing of information via a learning factory is an important area. However, since there is no direct contribution to production, this topic is also excluded from our research. Yield Forecasting increasingly uses techniques and concepts that are also applied in Cognitive Production Systems. Since these are processes in agriculture, these papers did not find their way into this work.

In addition to further sorting out of papers, the process also resulted in some questions, ambiguities and noteworthy secondary information within the included paper. IoT and Cloud/Edge Computing are mentioned and used in a large number of the underlying papers and form one of the main foundations to achieve Cognitive Manufacturing, but are not seen here as a concept or method. Although it is known that sustainability is an important optimization of many production systems and we consider it important for our topic "cognitive production systems", it is rarely mentioned in the papers and therefore has not found a way to answer the research questions. We see it as a positive side effect of optimizations such as "cost reduction" or "efficiency gain". 
The topic of ontologies was also covered by a total of eight papers. Generally speaking, machines are taught logical reasoning by mapping information. There is a great similarity to the topic of Patented Intelligence. In this topic the question remains open whether it is a concept in the sense of this paper or a method that is then implemented within a concept. It can also just be a help, with which one can then implement another specific method. In future work this open question can be clarified. Another important point is "cultural differences" in human-machine interfaces. For example, Chinese workers behave differently than German workers. Germans like direct access to simple functionality as uncomplicated as possible. Chinese, on the other hand, tend to press many buttons and prefer complicated, small-scale UI. This aspect has not been addressed in this paper, but it is also an important topic for an efficient use of human-machine interaction. 
There were still a large number of keywords in the individual papers, which are not otherwise discussed further, but which may be worth a closer isolated consideration in the future. Because these words occurred too seldom, they did not get a single category or further consideration. Among these keywords are:
\begin{itemize}
\item reduced intelligence
\item digital manufacturing
\item cognitive robots
\item cobots (cooperative robots)
\item responsive production
\end{itemize} 
Possible extensions can also be derived from the research questions and findings considered. One could still map the problems and limitations from research question 4 to the categories of research question 1 to see which problems occur most frequently in which category. From this one could derive which topic of which category one is dealing with in the future to solve this problem. New questions can be generated from the categorized problems of the papers. These questions could then be used to address exactly one of these problems. For example, the problem "too specific" could be addressed with a more general model that tries to generalize problems solved too specifically. Other research questions could be concerned with solving "non-functional requirements".

\section{conclusion and future work}\label{sec:conclusion-future-work}
We carried out extensive research on the topic of Cognitive Production Systems (CPS) in five scientific online databases. A first set of results consisting of 1255 papers from 2015-2019 was narrowed down to a final set of 88 papers. The carefully selected papers, corresponding to only 7.01 \% of the first result set, represent a representative amount of keywords documented above during this period. Four research questions were set up, which were then answered step by step by means of an analysis according to aspects such as categorisation, condition, advantages and problems. The established categories reflect different aspects of CPS such as human-machine interaction and multi-agent systems. The state of the papers in terms of theory and practice produced different results in the individual categories. For example, the category Knowledge Gaining/Sharing was significantly more practice-oriented than the other categories, while the category Human-Machine Interaction was evenly distributed across all states. The advantages identified partly coincided with the expectations of CPS. For example, the most frequently mentioned advantage "efficiency gain/improvement in effectiveness" best contributes to "mass customization" and the resulting shorter reaction time to changes. Other optimizations such as "information retrieval/distribution/use optimized" and "manual intervention reduced" also meet the expectations of modern production systems and contribute to better integration of people in production and better interaction with machines. Problems and barriers arise above all in the generalizability of the concepts discussed. Furthermore some approaches are too complex, too slow or too imprecise. However, the authors of the papers often mention plans to tackle the problems mentioned in the future and thus to further develop their concepts.

Generalizable concepts that can be used for various applications in production should be further researched in the future. Especially the use of Artificial Intelligence to automate complex tasks is worthwhile. At the same time, human beings and their role in production should also be considered. Cognitive skills should develop the human being into a supervisor rather than a mere production worker. Further research is also needed into solutions that provide humans with information in a compressed form that enables them to react quickly to the role of the supervisor even when changes occur frequently. Although the human being is still an important part of the production after evaluation of the available work, direct manual interventions in the production should be further reduced. It remains to be seen how long humans will be able to withstand the ever-increasing demands in terms of reaction speed, efficiency and amount of information, until they will eventually be completely replaced by machines in the production process. 

\bibliographystyle{ACM-Reference-Format}


\begin{thebibliography}{41}


\ifx \showCODEN    \undefined \def \showCODEN     #1{\unskip}     \fi
\ifx \showDOI      \undefined \def \showDOI       #1{#1}\fi
\ifx \showISBNx    \undefined \def \showISBNx     #1{\unskip}     \fi
\ifx \showISBNxiii \undefined \def \showISBNxiii  #1{\unskip}     \fi
\ifx \showISSN     \undefined \def \showISSN      #1{\unskip}     \fi
\ifx \showLCCN     \undefined \def \showLCCN      #1{\unskip}     \fi
\ifx \shownote     \undefined \def \shownote      #1{#1}          \fi
\ifx \showarticletitle \undefined \def \showarticletitle #1{#1}   \fi
\ifx \showURL      \undefined \def \showURL       {\relax}        \fi
\providecommand\bibfield[2]{#2}
\providecommand\bibinfo[2]{#2}
\providecommand\natexlab[1]{#1}
\providecommand\showeprint[2][]{arXiv:#2}

\bibitem[\protect\citeauthoryear{{Alessandro Roncone, Olivier Mangin, Brian
  Scassellati}}{{Alessandro Roncone, Olivier Mangin, Brian
  Scassellati}}{2017}]%
        {AlessandroRonconeOlivierManginBrianScassellati.2017}
\bibfield{author}{\bibinfo{person}{{Alessandro Roncone, Olivier Mangin, Brian
  Scassellati}}.} \bibinfo{year}{2017}\natexlab{}.
\newblock \bibinfo{booktitle}{\emph{Transparent Role Assignment and Task
  Allocation in Human Robot Collaboration: May 29-June 3, 2017, Singapore :
  ICRA 2017}}.
\newblock \bibinfo{publisher}{IEEE}, \bibinfo{address}{Piscataway, NJ}.
\newblock
\showISBNx{9781509046348}
\urldef\tempurl%
\url{http://ieeexplore.ieee.org/servlet/opac?punumber=7960754}
\showURL{%
\tempurl}


\bibitem[\protect\citeauthoryear{Alm, Aehnelt, and Urban}{Alm
  et~al\mbox{.}}{2015}]%
        {Alm.2015}
\bibfield{author}{\bibinfo{person}{Rebekka Alm}, \bibinfo{person}{Mario
  Aehnelt}, {and} \bibinfo{person}{Bodo Urban}.}
  \bibinfo{year}{2015}\natexlab{}.
\newblock \showarticletitle{Processing manufacturing knowledge with
  ontology-based annotations and cognitive architectures}. In
  \bibinfo{booktitle}{\emph{Proceedings of the 15th International Conference on
  Knowledge Technologies and Data-driven Business - i-KNOW '15}},
  \bibfield{editor}{\bibinfo{person}{Stefanie Lindstaedt},
  \bibinfo{person}{Tobias Ley}, {and} \bibinfo{person}{Harald Sack}} (Eds.).
  \bibinfo{publisher}{{ACM Press}}, \bibinfo{address}{New York, New York, USA},
  \bibinfo{pages}{1--6}.
\newblock
\showISBNx{9781450337212}
\urldef\tempurl%
\url{https://doi.org/10.1145/2809563.2809576}
\showDOI{\tempurl}


\bibitem[\protect\citeauthoryear{{Amedeo Cesta, Andrea Orlandini, Giulio
  Bernardi, Alessandro Umbrico$\dagger$}}{{Amedeo Cesta, Andrea Orlandini,
  Giulio Bernardi, Alessandro Umbrico$\dagger$}}{2016}]%
        {AmedeoCestaAndreaOrlandiniGiulioBernardiAlessandroUmbrico.2016}
\bibfield{author}{\bibinfo{person}{{Amedeo Cesta, Andrea Orlandini, Giulio
  Bernardi, Alessandro Umbrico$\dagger$}}.} \bibinfo{year}{2016}\natexlab{}.
\newblock \bibinfo{booktitle}{\emph{Towards a Planning-based Framework for
  Symbiotic Human-Robot Collaboration: September 6-9, 2016 Berlin, Germany}}.
\newblock \bibinfo{publisher}{IEEE}, \bibinfo{address}{Piscataway, NJ}.
\newblock
\showISBNx{9781509013159}
\urldef\tempurl%
\url{http://ieeexplore.ieee.org/xpl/mostRecentIssue.jsp?punumber=7593665}
\showURL{%
\tempurl}


\bibitem[\protect\citeauthoryear{Askarpour, Mandrioli, Rossi, and
  Vicentini}{Askarpour et~al\mbox{.}}{2017}]%
        {Askarpour.2017}
\bibfield{author}{\bibinfo{person}{Mehrnoosh Askarpour}, \bibinfo{person}{Dino
  Mandrioli}, \bibinfo{person}{Matteo Rossi}, {and} \bibinfo{person}{Federico
  Vicentini}.} \bibinfo{year}{2017}\natexlab{}.
\newblock \bibinfo{booktitle}{\emph{Modeling Operator Behavior in the Safety
  Analysis of Collaborative Robotic Applications}}.
  Vol.~\bibinfo{volume}{10488}.
\newblock \bibinfo{publisher}{{Springer International Publishing}},
  \bibinfo{address}{Cham}.
\newblock
\showISBNx{978-3-319-66265-7}
\urldef\tempurl%
\url{https://doi.org/10.1007/978-3-319-66266-4}
\showDOI{\tempurl}


\bibitem[\protect\citeauthoryear{Bolmsj{\"o}}{Bolmsj{\"o}}{2015}]%
        {Bolmsjo.2015}
\bibfield{author}{\bibinfo{person}{Gunnar Bolmsj{\"o}}.}
  \bibinfo{year}{2015}\natexlab{}.
\newblock \showarticletitle{Supporting Tools for Operator in Robot
  Collaborative Mode}.
\newblock \bibinfo{journal}{\emph{Procedia Manufacturing}}  \bibinfo{volume}{3}
  (\bibinfo{year}{2015}), \bibinfo{pages}{409--416}.
\newblock
\showISSN{23519789}
\urldef\tempurl%
\url{https://doi.org/10.1016/j.promfg.2015.07.190}
\showDOI{\tempurl}


\bibitem[\protect\citeauthoryear{Borisov, Borisov, Gromov, Vlasov, and
  Kolyubin}{Borisov et~al\mbox{.}}{2018}]%
        {Borisov.2018}
\bibfield{author}{\bibinfo{person}{Ivan~I. Borisov}, \bibinfo{person}{Oleg~I.
  Borisov}, \bibinfo{person}{Vladislav~S. Gromov}, \bibinfo{person}{Sergey~M.
  Vlasov}, {and} \bibinfo{person}{Sergey~A. Kolyubin}.}
  \bibinfo{year}{15.05.2018 - 18.05.2018}\natexlab{}.
\newblock \showarticletitle{Versatile Gripper as Key Part For Smart Factory}.
  In \bibinfo{booktitle}{\emph{2018 IEEE Industrial Cyber-Physical Systems
  (ICPS)}}. \bibinfo{publisher}{IEEE}, \bibinfo{pages}{476--481}.
\newblock
\showISBNx{978-1-5386-6531-2}
\urldef\tempurl%
\url{https://doi.org/10.1109/ICPHYS.2018.8390751}
\showDOI{\tempurl}


\bibitem[\protect\citeauthoryear{Breivold}{Breivold}{2017}]%
        {Breivold.2017}
\bibfield{author}{\bibinfo{person}{Hongyu~Pei Breivold}.}
  \bibinfo{year}{2017}\natexlab{}.
\newblock \showarticletitle{Internet-of-things and Cloud Computing for Smart
  Industry: A Systematic Mapping Study}. In \bibinfo{booktitle}{\emph{2017 5TH
  INTERNATIONAL CONFERENCE ON ENTERPRISE SYSTEMS (ES)}}
  \emph{(\bibinfo{series}{International Conference on Enterprise Systems
  (ES)})}, \bibfield{editor}{\bibinfo{person}{Z.~Pang},
  \bibinfo{person}{L.~Li}, {and} \bibinfo{person}{G.~Li}} (Eds.).
  \bibinfo{publisher}{{IEEE COMPUTER SOC}}, \bibinfo{address}{10662 LOS
  VAQUEROS CIRCLE, PO BOX 3014, LOS ALAMITOS, CA 90720-1264 USA},
  \bibinfo{pages}{299--304}.
\newblock
\showISBNx{978-1-5386-0936-1}
\urldef\tempurl%
\url{https://doi.org/10.1109/ES.2017.56}
\showDOI{\tempurl}


\bibitem[\protect\citeauthoryear{Coupet{\'e}, Moutarde, and
  Manitsaris}{Coupet{\'e} et~al\mbox{.}}{2016}]%
        {Coupete.2016}
\bibfield{author}{\bibinfo{person}{Eva Coupet{\'e}}, \bibinfo{person}{Fabien
  Moutarde}, {and} \bibinfo{person}{Sotiris Manitsaris}.}
  \bibinfo{year}{2016}\natexlab{}.
\newblock \showarticletitle{A User-Adaptive Gesture Recognition System Applied
  to Human-Robot Collaboration in Factories}. In
  \bibinfo{booktitle}{\emph{Proceedings of the 3rd International Symposium on
  Movement and Computing - MOCO '16}},
  \bibfield{editor}{\bibinfo{person}{Sotiris Manitsaris}} (Ed.).
  \bibinfo{publisher}{{ACM Press}}, \bibinfo{address}{New York, New York, USA},
  \bibinfo{pages}{1--7}.
\newblock
\showISBNx{9781450343077}
\urldef\tempurl%
\url{https://doi.org/10.1145/2948910.2948933}
\showDOI{\tempurl}


\bibitem[\protect\citeauthoryear{Datta and Bonnet}{Datta and Bonnet}{2018}]%
        {Soumya.2018}
\bibfield{author}{\bibinfo{person}{Soumya~Kanti Datta} {and}
  \bibinfo{person}{Christian Bonnet}.} \bibinfo{year}{16.12.2018 -
  19.12.2018}\natexlab{}.
\newblock \showarticletitle{MEC and IoT Based Automatic Agent Reconfiguration
  in Industry 4.0}. In \bibinfo{booktitle}{\emph{2018 IEEE International
  Conference on Advanced Networks and Telecommunications Systems (ANTS)}}.
  \bibinfo{publisher}{IEEE}, \bibinfo{pages}{1--5}.
\newblock
\showISBNx{978-1-5386-8134-3}
\urldef\tempurl%
\url{https://doi.org/10.1109/ANTS.2018.8710126}
\showDOI{\tempurl}


\bibitem[\protect\citeauthoryear{{David Romero, Johan Stahre, Thorsten Wuest,
  Ovidiu Noran, Peter Bernus, {\AA}sa Fast-Berglund, Dominic Gorecky}}{{David
  Romero, Johan Stahre, Thorsten Wuest, Ovidiu Noran, Peter Bernus, {\AA}sa
  Fast-Berglund, Dominic Gorecky}}{[n.d.]}]%
  {DavidRomeroJohanStahreThorstenWuestOvidiuNoranPeterBernusAsaFastBerglundDominic.}
\bibfield{author}{\bibinfo{person}{{David Romero, Johan Stahre, Thorsten Wuest,
  Ovidiu Noran, Peter Bernus, {\AA}sa Fast-Berglund, Dominic Gorecky}}.}
  \bibinfo{year}{[n.d.]}\natexlab{}.
\newblock \showarticletitle{TOWARDS AN OPERATOR 4.0 TYPOLOGY: A HUMAN-CENTRIC
  PERSPECTIVE ON THE FOURTH INDUSTRIAL REVOLUTION TECHNOLOGIES}.
\newblock  (\bibinfo{year}{[n.\,d.]}).
\newblock


\bibitem[\protect\citeauthoryear{Dey and Sarkar}{Dey and Sarkar}{2016}]%
        {Dey.2016}
\bibfield{author}{\bibinfo{person}{Surojit Dey} {and} \bibinfo{person}{Pratiti
  Sarkar}.} \bibinfo{year}{2016}\natexlab{}.
\newblock \showarticletitle{Augmented Reality Based Integrated Intelligent
  Maintenance System For Production Line}. In
  \bibinfo{booktitle}{\emph{Proceedings of the 8th Indian Conference on Human
  Computer Interaction - IHCI '16}},
  \bibfield{editor}{\bibinfo{person}{Unknown}} (Ed.). \bibinfo{publisher}{{ACM
  Press}}, \bibinfo{address}{New York, New York, USA},
  \bibinfo{pages}{126--131}.
\newblock
\showISBNx{9781450348638}
\urldef\tempurl%
\url{https://doi.org/10.1145/3014362.3014377}
\showDOI{\tempurl}


\bibitem[\protect\citeauthoryear{Duguay, Landry, and Pasin}{Duguay
  et~al\mbox{.}}{1997}]%
        {Duguay.1997}
\bibfield{author}{\bibinfo{person}{Claude~R. Duguay}, \bibinfo{person}{Sylvain
  Landry}, {and} \bibinfo{person}{Federico Pasin}.}
  \bibinfo{year}{1997}\natexlab{}.
\newblock \showarticletitle{From mass production to flexible/agile production}.
\newblock \bibinfo{journal}{\emph{International Journal of Operations {\&}
  Production Management}} \bibinfo{volume}{17}, \bibinfo{number}{12}
  (\bibinfo{year}{1997}), \bibinfo{pages}{1183--1195}.
\newblock
\showISSN{0144-3577}
\urldef\tempurl%
\url{https://doi.org/10.1108/01443579710182936}
\showDOI{\tempurl}


\bibitem[\protect\citeauthoryear{Emmanouilidis, Pistofidis, Bertoncelj,
  Katsouros, Fournaris, Koulamas, and Ruiz-Carcel}{Emmanouilidis
  et~al\mbox{.}}{2019}]%
        {Emmanouilidis.2019}
\bibfield{author}{\bibinfo{person}{Christos Emmanouilidis},
  \bibinfo{person}{Petros Pistofidis}, \bibinfo{person}{Luka Bertoncelj},
  \bibinfo{person}{Vassilis Katsouros}, \bibinfo{person}{Apostolos Fournaris},
  \bibinfo{person}{Christos Koulamas}, {and} \bibinfo{person}{Cristobal
  Ruiz-Carcel}.} \bibinfo{year}{2019}\natexlab{}.
\newblock \showarticletitle{Enabling the human in the loop: Linked data and
  knowledge in industrial cyber-physical systems}.
\newblock \bibinfo{journal}{\emph{Annual Reviews in Control}}
  \bibinfo{volume}{47} (\bibinfo{year}{2019}), \bibinfo{pages}{249--265}.
\newblock
\showISSN{1367-5788}
\urldef\tempurl%
\url{https://doi.org/10.1016/j.arcontrol.2019.03.004}
\showDOI{\tempurl}


\bibitem[\protect\citeauthoryear{Goebel, Chander, Holzinger, Lecue, Akata,
  Stumpf, Kieseberg, and Holzinger}{Goebel et~al\mbox{.}}{2018}]%
        {Goebel.2018}
\bibfield{author}{\bibinfo{person}{Randy Goebel}, \bibinfo{person}{Ajay
  Chander}, \bibinfo{person}{Katharina Holzinger}, \bibinfo{person}{Freddy
  Lecue}, \bibinfo{person}{Zeynep Akata}, \bibinfo{person}{Simone Stumpf},
  \bibinfo{person}{Peter Kieseberg}, {and} \bibinfo{person}{Andreas
  Holzinger}.} \bibinfo{year}{2018}\natexlab{}.
\newblock \showarticletitle{Explainable AI: The New 42?}
\newblock In \bibinfo{booktitle}{\emph{Machine Learning and Knowledge
  Extraction}}, \bibfield{editor}{\bibinfo{person}{Andreas Holzinger},
  \bibinfo{person}{Peter Kieseberg}, \bibinfo{person}{A.~Min Tjoa}, {and}
  \bibinfo{person}{Edgar Weippl}} (Eds.). \bibinfo{series}{Lecture Notes in
  Computer Science}, Vol.~\bibinfo{volume}{11015}.
  \bibinfo{publisher}{{Springer International Publishing}},
  \bibinfo{address}{Cham}, \bibinfo{pages}{295--303}.
\newblock
\showISBNx{978-3-319-99739-1}
\urldef\tempurl%
\url{https://doi.org/10.1007/978-3-319-99740-7{\textunderscore }21}
\showDOI{\tempurl}


\bibitem[\protect\citeauthoryear{{HyeongGon Jo}, {Hyo Jeon Kwon}, and {Jae Duck
  Lee}}{{HyeongGon Jo} et~al\mbox{.}}{2017}]%
        {HyeongGonJo.2017}
\bibfield{author}{\bibinfo{person}{SoonJu~Kang {HyeongGon Jo}},
  \bibinfo{person}{{Hyo Jeon Kwon}}, {and} \bibinfo{person}{{Jae Duck Lee}}.}
  \bibinfo{year}{2017}\natexlab{}.
\newblock \bibinfo{booktitle}{\emph{In-door Location-based Smart Factory Cloud
  Platform supporting Device-to-Device Self-Collaboration: 13-16 Feb. 2017}}.
\newblock \bibinfo{publisher}{IEEE}, \bibinfo{address}{Piscataway, NJ}.
\newblock
\showISBNx{9781509030163}
\urldef\tempurl%
\url{http://ieeexplore.ieee.org/servlet/opac?punumber=7877084}
\showURL{%
\tempurl}


\bibitem[\protect\citeauthoryear{{J. Jokinen} and {J. L. M. Lastra}}{{J.
  Jokinen} and {J. L. M. Lastra}}{2016}]%
        {J.Jokinen.2016}
\bibfield{author}{\bibinfo{person}{{J. Jokinen}} {and} \bibinfo{person}{{J. L.
  M. Lastra}}.} \bibinfo{year}{2016}\natexlab{}.
\newblock \bibinfo{booktitle}{\emph{Industrial monitoring and control approach
  for dynamic and distributed intelligent systems: September 6-9, 2016 Berlin,
  Germany}}.
\newblock \bibinfo{publisher}{IEEE}, \bibinfo{address}{Piscataway, NJ}.
\newblock
\showISBNx{9781509013159}
\urldef\tempurl%
\url{http://ieeexplore.ieee.org/xpl/mostRecentIssue.jsp?punumber=7593665}
\showURL{%
\tempurl}


\bibitem[\protect\citeauthoryear{Jost and S{\"u}sser}{Jost and
  S{\"u}sser}{2020}]%
        {Jost.2020}
\bibfield{author}{\bibinfo{person}{Peter-J. Jost} {and}
  \bibinfo{person}{Theresa S{\"u}sser}.} \bibinfo{year}{2020}\natexlab{}.
\newblock \showarticletitle{Company-customer interaction in mass
  customization}.
\newblock \bibinfo{journal}{\emph{International Journal of Production
  Economics}}  \bibinfo{volume}{220} (\bibinfo{year}{2020}),
  \bibinfo{pages}{107454}.
\newblock
\showISSN{09255273}
\urldef\tempurl%
\url{https://doi.org/10.1016/j.ijpe.2019.07.027}
\showDOI{\tempurl}


\bibitem[\protect\citeauthoryear{Lee, Ryu, and Shin}{Lee et~al\mbox{.}}{2017}]%
        {Lee.2017}
\bibfield{author}{\bibinfo{person}{Sangil Lee}, \bibinfo{person}{Kwangyeol
  Ryu}, {and} \bibinfo{person}{Moonsoo Shin}.} \bibinfo{year}{2017}\natexlab{}.
\newblock \showarticletitle{The Development of Simulation Model for
  Self-reconfigurable Manufacturing System Considering Sustainability Factors}.
\newblock \bibinfo{journal}{\emph{Procedia Manufacturing}}
  \bibinfo{volume}{11} (\bibinfo{year}{2017}), \bibinfo{pages}{1085--1092}.
\newblock
\showISSN{23519789}
\urldef\tempurl%
\url{https://doi.org/10.1016/j.promfg.2017.07.226}
\showDOI{\tempurl}


\bibitem[\protect\citeauthoryear{Mabkhot, Al-Samhan, and Darmoul}{Mabkhot
  et~al\mbox{.}}{2016}]%
        {Mabkhot.2016}
\bibfield{author}{\bibinfo{person}{Mohammed.~M. Mabkhot},
  \bibinfo{person}{Ali~M. Al-Samhan}, {and} \bibinfo{person}{Saber Darmoul}.}
  \bibinfo{year}{2016}\natexlab{}.
\newblock \showarticletitle{An information model to support reconfiguration of
  manufacturing systems}.
\newblock \bibinfo{journal}{\emph{IFAC-PapersOnLine}} \bibinfo{volume}{49},
  \bibinfo{number}{5} (\bibinfo{year}{2016}), \bibinfo{pages}{37--42}.
\newblock
\showISSN{24058963}
\urldef\tempurl%
\url{https://doi.org/10.1016/j.ifacol.2016.07.086}
\showDOI{\tempurl}


\bibitem[\protect\citeauthoryear{Mayer and Michahelles}{Mayer and
  Michahelles}{2017}]%
        {Mayer.2017}
\bibfield{author}{\bibinfo{person}{Simon Mayer} {and} \bibinfo{person}{Florian
  Michahelles}.} \bibinfo{year}{2017}\natexlab{}.
\newblock \showarticletitle{HoloInteractions: Visualizing Interactions Between
  Autonomous Cognitive Machines}. In \bibinfo{booktitle}{\emph{Proceedings of
  the Seventh International Conference on the Internet of Things}}
  \emph{(\bibinfo{series}{IoT '17})}. \bibinfo{publisher}{ACM},
  \bibinfo{address}{New York, NY, USA}, \bibinfo{pages}{40:1--40:2}.
\newblock
\showISBNx{978-1-4503-5318-2}
\urldef\tempurl%
\url{https://doi.org/10.1145/3131542.3140278}
\showDOI{\tempurl}


\bibitem[\protect\citeauthoryear{Moghaddam, Cadavid, Kenley, and
  Deshmukh}{Moghaddam et~al\mbox{.}}{2018}]%
        {Moghaddam.2018}
\bibfield{author}{\bibinfo{person}{Mohsen Moghaddam},
  \bibinfo{person}{Marissa~N. Cadavid}, \bibinfo{person}{C.~Robert Kenley},
  {and} \bibinfo{person}{Abhijit~V. Deshmukh}.}
  \bibinfo{year}{2018}\natexlab{}.
\newblock \showarticletitle{Reference architectures for smart manufacturing: A
  critical review}.
\newblock \bibinfo{journal}{\emph{Journal of Manufacturing Systems}}
  \bibinfo{volume}{49} (\bibinfo{year}{2018}), \bibinfo{pages}{215--225}.
\newblock
\urldef\tempurl%
\url{https://doi.org/10.1016/j.jmsy.2018.10.006}
\showDOI{\tempurl}


\bibitem[\protect\citeauthoryear{Murauer, Gerhard, Gollan, Haslgr{\"u}bler,
  Selymes, Sopidis, St{\"u}tz, Wei{\ss}enbach, Jungwirth, Anzengruber, Abbas,
  Ahmad, Azadi, Cho, Ennsbrunner, and Ferscha}{Murauer et~al\mbox{.}}{2019}]%
        {Murauer.2019}
\bibfield{author}{\bibinfo{person}{Michaela Murauer}, \bibinfo{person}{Detlef
  Gerhard}, \bibinfo{person}{Benedikt Gollan}, \bibinfo{person}{Michael
  Haslgr{\"u}bler}, \bibinfo{person}{Johannes Selymes},
  \bibinfo{person}{Georgios Sopidis}, \bibinfo{person}{Matthias St{\"u}tz},
  \bibinfo{person}{Paul Wei{\ss}enbach}, \bibinfo{person}{Florian Jungwirth},
  \bibinfo{person}{Bernhard Anzengruber}, \bibinfo{person}{Ali Abbas},
  \bibinfo{person}{Abdelrahman Ahmad}, \bibinfo{person}{Behrooz Azadi},
  \bibinfo{person}{Jullius Cho}, \bibinfo{person}{Helmut Ennsbrunner}, {and}
  \bibinfo{person}{Alois Ferscha}.} \bibinfo{year}{2019}\natexlab{}.
\newblock \showarticletitle{A task-independent design and development process
  for cognitive products in industrial applications}. In
  \bibinfo{booktitle}{\emph{Proceedings of the 12th ACM International
  Conference on PErvasive Technologies Related to Assistive Environments -
  PETRA '19}}, \bibfield{editor}{\bibinfo{person}{Fillia Makedon}} (Ed.).
  \bibinfo{publisher}{{ACM Press}}, \bibinfo{address}{New York, New York, USA},
  \bibinfo{pages}{358--367}.
\newblock
\showISBNx{9781450362320}
\urldef\tempurl%
\url{https://doi.org/10.1145/3316782.3322748}
\showDOI{\tempurl}


\bibitem[\protect\citeauthoryear{{Numan M. Durakbasa} and {G{\"o}kcen
  Bas}}{{Numan M. Durakbasa} and {G{\"o}kcen Bas}}{[n.d.]}]%
        {NumanM.Durakbasa.}
\bibfield{author}{\bibinfo{person}{{Numan M. Durakbasa}} {and}
  \bibinfo{person}{{G{\"o}kcen Bas}}.} \bibinfo{year}{[n.d.]}\natexlab{}.
\newblock \showarticletitle{From Nano- to Pico- to Femto-: Future Challenges
  for Cost-Oriented Manufacturing}.
\newblock  (\bibinfo{year}{[n.\,d.]}).
\newblock


\bibitem[\protect\citeauthoryear{{Olga Meyer}, {Greg Rauhoeft}, {Daniel Schel},
  and {Daniel Stock}}{{Olga Meyer} et~al\mbox{.}}{2018}]%
        {OlgaMeyer.2018}
\bibfield{author}{\bibinfo{person}{{Olga Meyer}}, \bibinfo{person}{{Greg
  Rauhoeft}}, \bibinfo{person}{{Daniel Schel}}, {and} \bibinfo{person}{{Daniel
  Stock}}.} \bibinfo{year}{2018}\natexlab{}.
\newblock \showarticletitle{Industrial Internet of Things: covering
  standardization gaps for the next generation of reconfigurable production
  systems}.
\newblock  (\bibinfo{year}{2018}).
\newblock


\bibitem[\protect\citeauthoryear{Osterrieder, Budde, and Friedli}{Osterrieder
  et~al\mbox{.}}{2019}]%
        {Osterrieder.2019}
\bibfield{author}{\bibinfo{person}{Philipp Osterrieder}, \bibinfo{person}{Lukas
  Budde}, {and} \bibinfo{person}{Thomas Friedli}.}
  \bibinfo{year}{2019}\natexlab{}.
\newblock \showarticletitle{The smart factory as a key construct of industry
  4.0: A systematic literature review}.
\newblock \bibinfo{journal}{\emph{International Journal of Production
  Economics}} (\bibinfo{year}{2019}).
\newblock
\showISSN{09255273}
\urldef\tempurl%
\url{https://doi.org/10.1016/j.ijpe.2019.08.011}
\showDOI{\tempurl}


\bibitem[\protect\citeauthoryear{Pacaux-Lemoine, Berdal, Enjalbert, and
  Trentesaux}{Pacaux-Lemoine et~al\mbox{.}}{2018}]%
        {PacauxLemoine.15.05.201818.05.2018}
\bibfield{author}{\bibinfo{person}{Marie-Pierre Pacaux-Lemoine},
  \bibinfo{person}{Quentin Berdal}, \bibinfo{person}{Simon Enjalbert}, {and}
  \bibinfo{person}{Damien Trentesaux}.} \bibinfo{year}{15.05.2018 -
  18.05.2018}\natexlab{}.
\newblock \showarticletitle{Towards human-based industrial cyber-physical
  systems}. In \bibinfo{booktitle}{\emph{2018 IEEE Industrial Cyber-Physical
  Systems (ICPS)}}. \bibinfo{publisher}{IEEE}, \bibinfo{pages}{615--620}.
\newblock
\showISBNx{978-1-5386-6531-2}
\urldef\tempurl%
\url{https://doi.org/10.1109/ICPHYS.2018.8390776}
\showDOI{\tempurl}


\bibitem[\protect\citeauthoryear{Panetto, Iung, Ivanov, Weichhart, and
  Wang}{Panetto et~al\mbox{.}}{2019}]%
        {Panetto.2019}
\bibfield{author}{\bibinfo{person}{Herv{\'e} Panetto}, \bibinfo{person}{Benoit
  Iung}, \bibinfo{person}{Dmitry Ivanov}, \bibinfo{person}{Georg Weichhart},
  {and} \bibinfo{person}{Xiaofan Wang}.} \bibinfo{year}{2019}\natexlab{}.
\newblock \showarticletitle{Challenges for the cyber-physical manufacturing
  enterprises of the future}.
\newblock \bibinfo{journal}{\emph{Annual Reviews in Control}}
  \bibinfo{volume}{47} (\bibinfo{year}{2019}), \bibinfo{pages}{200--213}.
\newblock
\showISSN{1367-5788}
\urldef\tempurl%
\url{https://doi.org/10.1016/j.arcontrol.2019.02.002}
\showDOI{\tempurl}


\bibitem[\protect\citeauthoryear{Peruzzini, Pellicciari, and Bil}{Peruzzini
  et~al\mbox{.}}{2018}]%
        {Peruzzini.2018}
\bibfield{author}{\bibinfo{person}{M. Peruzzini}, \bibinfo{person}{M.
  Pellicciari}, {and} \bibinfo{person}{C. Bil}.}
  \bibinfo{year}{2018}\natexlab{}.
\newblock \bibinfo{booktitle}{\emph{Adaptive Inspection Cell for HMI
  Consoles}}. \bibinfo{series}{Advances in Transdisciplinary Engineering Ser},
  Vol.~\bibinfo{volume}{v.7}.
\newblock \bibinfo{publisher}{{IOS Press Incorporated}},
  \bibinfo{address}{Amsterdam}.
\newblock
\showISBNx{9781614998976}
\urldef\tempurl%
\url{https://ebookcentral.proquest.com/lib/gbv/detail.action?docID=5543857}
\showURL{%
\tempurl}


\bibitem[\protect\citeauthoryear{{Pfeifer, Schmitt, Stemmer, Rollof, Schneider,
  Doro}}{{Pfeifer, Schmitt, Stemmer, Rollof, Schneider, Doro}}{2010}]%
        {PfeiferSchmittStemmerRollofSchneiderDoro.2010}
\bibfield{author}{\bibinfo{person}{{Pfeifer, Schmitt, Stemmer, Rollof,
  Schneider, Doro}}.} \bibinfo{year}{2010}\natexlab{}.
\newblock \showarticletitle{Cognitive Production Metrology: A new concept for
  flexibly attending the inspection requirements of small series production}.
\newblock  (\bibinfo{year}{2010}).
\newblock


\bibitem[\protect\citeauthoryear{Raj, Dwivedi, Sharma, {Lopes de Sousa
  Jabbour}, and Rajak}{Raj et~al\mbox{.}}{2019}]%
        {Raj.2019}
\bibfield{author}{\bibinfo{person}{Alok Raj}, \bibinfo{person}{Gourav Dwivedi},
  \bibinfo{person}{Ankit Sharma}, \bibinfo{person}{Ana~Beatriz {Lopes de Sousa
  Jabbour}}, {and} \bibinfo{person}{Sonu Rajak}.}
  \bibinfo{year}{2019}\natexlab{}.
\newblock \showarticletitle{Barriers to the adoption of industry 4.0
  technologies in the manufacturing sector: An inter-country comparative
  perspective}.
\newblock \bibinfo{journal}{\emph{International Journal of Production
  Economics}} (\bibinfo{year}{2019}), \bibinfo{pages}{107546}.
\newblock
\showISSN{09255273}
\urldef\tempurl%
\url{https://doi.org/10.1016/j.ijpe.2019.107546}
\showDOI{\tempurl}


\bibitem[\protect\citeauthoryear{Ruppert, Jask{\'o}, Holczinger, and
  Abonyi}{Ruppert et~al\mbox{.}}{2018}]%
        {Ruppert.2018}
\bibfield{author}{\bibinfo{person}{Tam{\'a}s Ruppert},
  \bibinfo{person}{Szil{\'a}rd Jask{\'o}}, \bibinfo{person}{Tibor Holczinger},
  {and} \bibinfo{person}{J{\'a}nos Abonyi}.} \bibinfo{year}{2018}\natexlab{}.
\newblock \showarticletitle{Enabling Technologies for Operator 4.0: A Survey}.
\newblock \bibinfo{journal}{\emph{Applied Sciences}} \bibinfo{volume}{8},
  \bibinfo{number}{9} (\bibinfo{year}{2018}), \bibinfo{pages}{1650}.
\newblock
\urldef\tempurl%
\url{https://doi.org/10.3390/app8091650}
\showDOI{\tempurl}


\bibitem[\protect\citeauthoryear{{S. M. Mizanoor Rahman}}{{S. M. Mizanoor
  Rahman}}{[n.d.]}]%
        {S.M.MizanoorRahman.}
\bibfield{author}{\bibinfo{person}{{S. M. Mizanoor Rahman}}.}
  \bibinfo{year}{[n.d.]}\natexlab{}.
\newblock \showarticletitle{Cognitive Cyber-Physical System (C-CPS) for
  Human-Robot Collaborative Manufacturing}.
\newblock  (\bibinfo{year}{[n.\,d.]}).
\newblock
\urldef\tempurl%
\url{https://www2.scopus.com/record/display.uri?eid=2-s2.0-85069759622&origin=resultslist&sort=plf-f&src=s&st1=cognitive+manufacturing&nlo=&nlr=&nls=&sid=3f942fab0e9a7bf9aabacac84d4d906a&sot=b&sdt=cl&cluster=scosubjabbr\%2c"SOCI"\%2cf\%2bscosubjabbr\%2c"MEDI"\%2cf\%2c"PSYC"\%2cf&sl=38&s=TITLE-ABS-KEY\%28cognitive+manufacturing\%29&relpos=25&citeCnt=0&searchTerm=}
\showURL{%
\tempurl}


\bibitem[\protect\citeauthoryear{Sadik, Urban, and Adel}{Sadik
  et~al\mbox{.}}{2017}]%
        {Sadik.2017}
\bibfield{author}{\bibinfo{person}{Ahmed~R. Sadik}, \bibinfo{person}{Bodo
  Urban}, {and} \bibinfo{person}{Omar Adel}.} \bibinfo{year}{2017}\natexlab{}.
\newblock \showarticletitle{Using Hand Gestures to Interact with an Industrial
  Robot in a Cooperative Flexible Manufacturing Scenario}. In
  \bibinfo{booktitle}{\emph{Proceedings of the 3rd International Conference on
  Mechatronics and Robotics Engineering - ICMRE 2017}},
  \bibfield{editor}{\bibinfo{person}{Unknown}} (Ed.). \bibinfo{publisher}{{ACM
  Press}}, \bibinfo{address}{New York, New York, USA}, \bibinfo{pages}{11--16}.
\newblock
\showISBNx{9781450352802}
\urldef\tempurl%
\url{https://doi.org/10.1145/3068796.3068801}
\showDOI{\tempurl}


\bibitem[\protect\citeauthoryear{Sarkar and {\v{S}}ormaz}{Sarkar and
  {\v{S}}ormaz}{2018}]%
        {Sarkar.2018}
\bibfield{author}{\bibinfo{person}{Arkopaul Sarkar} {and}
  \bibinfo{person}{Du{\v{s}}an {\v{S}}ormaz}.} \bibinfo{year}{2018}\natexlab{}.
\newblock \showarticletitle{Multi-agent System for Cloud Manufacturing Process
  Planning}.
\newblock \bibinfo{journal}{\emph{Procedia Manufacturing}}
  \bibinfo{volume}{17} (\bibinfo{year}{2018}), \bibinfo{pages}{435--443}.
\newblock
\showISSN{23519789}
\urldef\tempurl%
\url{https://doi.org/10.1016/j.promfg.2018.10.067}
\showDOI{\tempurl}


\bibitem[\protect\citeauthoryear{Sharp, Ak, and Hedberg}{Sharp
  et~al\mbox{.}}{2018}]%
        {Sharp.2018}
\bibfield{author}{\bibinfo{person}{Michael Sharp}, \bibinfo{person}{Ronay Ak},
  {and} \bibinfo{person}{Thomas Hedberg}.} \bibinfo{year}{2018}\natexlab{}.
\newblock \showarticletitle{A Survey of the Advancing Use and Development of
  Machine Learning in Smart Manufacturing}.
\newblock \bibinfo{journal}{\emph{Journal of Manufacturing Systems}}
  \bibinfo{volume}{48 Pt C} (\bibinfo{year}{2018}).
\newblock
\urldef\tempurl%
\url{https://doi.org/10.1016/j.jmsy.2018.02.004}
\showDOI{\tempurl}


\bibitem[\protect\citeauthoryear{{Tom Bewley} and {Minas Liarokapis}}{{Tom
  Bewley} and {Minas Liarokapis}}{2019}]%
        {TomBewley.2019}
\bibfield{author}{\bibinfo{person}{{Tom Bewley}} {and} \bibinfo{person}{{Minas
  Liarokapis}}.} \bibinfo{year}{2019}\natexlab{}.
\newblock \showarticletitle{On the Combination of Gamification and Crowd
  Computation in Industrial Automation and Robotics Applications}.
\newblock  (\bibinfo{year}{2019}).
\newblock


\bibitem[\protect\citeauthoryear{Tran, Park, Nguyen, and Hoang}{Tran
  et~al\mbox{.}}{2019}]%
        {Tran.2019}
\bibfield{author}{\bibinfo{person}{Tran}, \bibinfo{person}{Park},
  \bibinfo{person}{Nguyen}, {and} \bibinfo{person}{Hoang}.}
  \bibinfo{year}{2019}\natexlab{}.
\newblock \showarticletitle{Development of a Smart Cyber-Physical Manufacturing
  System in the Industry 4.0 Context}.
\newblock \bibinfo{journal}{\emph{Applied Sciences}} \bibinfo{volume}{9},
  \bibinfo{number}{16} (\bibinfo{year}{2019}), \bibinfo{pages}{3325}.
\newblock
\urldef\tempurl%
\url{https://doi.org/10.3390/app9163325}
\showDOI{\tempurl}


\bibitem[\protect\citeauthoryear{{Volker Krueger, Francesco Rovida, Bjarne
  Grossmann, Ronald Petrick, Mathew Crosby, Arnaud Charzoule, German Martin
  Garcia, Sven Bahnke, Cesar Toscano, Germano Veiga}}{{Volker Krueger,
  Francesco Rovida, Bjarne Grossmann, Ronald Petrick, Mathew Crosby, Arnaud
  Charzoule, German Martin Garcia, Sven Bahnke, Cesar Toscano, Germano
  Veiga}}{2019}]%
  {VolkerKruegerFrancescoRovidaBjarneGrossmannRonaldPetrickMathewCrosbyArnaudCharzoule.2019}
\bibfield{author}{\bibinfo{person}{{Volker Krueger, Francesco Rovida, Bjarne
  Grossmann, Ronald Petrick, Mathew Crosby, Arnaud Charzoule, German Martin
  Garcia, Sven Bahnke, Cesar Toscano, Germano Veiga}}.}
  \bibinfo{year}{2019}\natexlab{}.
\newblock \showarticletitle{Testing the vertical and cyber-physical integration
  of cognitive robots in manufacturing}.
\newblock \bibinfo{journal}{\emph{Robotics and Computer Integrated
  Manufacturing}} \bibinfo{number}{57} (\bibinfo{year}{2019}),
  \bibinfo{pages}{213--229}.
\newblock
\urldef\tempurl%
\url{https://www.deepdyve.com/lp/elsevier/testing-the-vertical-and-cyber-physical-integration-of-cognitive-GNecl7XKy0?key=elsevier}
\showURL{%
\tempurl}


\bibitem[\protect\citeauthoryear{Wan, Tang, {Di Li}, Imran, Zhang, Liu, and
  Pang}{Wan et~al\mbox{.}}{2019}]%
        {Wan.2019}
\bibfield{author}{\bibinfo{person}{Jiafu Wan}, \bibinfo{person}{Shenglong
  Tang}, \bibinfo{person}{{Di Li}}, \bibinfo{person}{Muhammad Imran},
  \bibinfo{person}{Chunhua Zhang}, \bibinfo{person}{Chengliang Liu}, {and}
  \bibinfo{person}{Zhibo Pang}.} \bibinfo{year}{2019}\natexlab{}.
\newblock \showarticletitle{Reconfigurable Smart Factory for Drug Packing in
  Healthcare Industry 4.0}.
\newblock \bibinfo{journal}{\emph{IEEE Transactions on Industrial Informatics}}
  \bibinfo{volume}{15}, \bibinfo{number}{1} (\bibinfo{year}{2019}),
  \bibinfo{pages}{507--516}.
\newblock
\showISSN{1551-3203}
\urldef\tempurl%
\url{https://doi.org/10.1109/TII.2018.2843811}
\showDOI{\tempurl}


\bibitem[\protect\citeauthoryear{Wang, Sun, Zhang, Thomas, Duan, and Shi}{Wang
  et~al\mbox{.}}{2016}]%
        {Wang.2016}
\bibfield{author}{\bibinfo{person}{JunPing Wang}, \bibinfo{person}{YunChuan
  Sun}, \bibinfo{person}{WenSheng Zhang}, \bibinfo{person}{Ian Thomas},
  \bibinfo{person}{ShiHui Duan}, {and} \bibinfo{person}{YouKang Shi}.}
  \bibinfo{year}{2016}\natexlab{}.
\newblock \showarticletitle{Large-Scale Online Multitask Learning and Decision
  Making for Flexible Manufacturing}.
\newblock \bibinfo{journal}{\emph{IEEE Transactions on Industrial Informatics}}
  \bibinfo{volume}{12}, \bibinfo{number}{6} (\bibinfo{year}{2016}),
  \bibinfo{pages}{2139--2147}.
\newblock
\showISSN{1551-3203}
\urldef\tempurl%
\url{https://doi.org/10.1109/TII.2016.2549919}
\showDOI{\tempurl}


\bibitem[\protect\citeauthoryear{Zolotov{\'a}, Papcun, Kaj{\'a}ti,
  Mi{\v{s}}kuf, and Mocnej}{Zolotov{\'a} et~al\mbox{.}}{2018}]%
        {Zolotova.2018}
\bibfield{author}{\bibinfo{person}{Iveta Zolotov{\'a}}, \bibinfo{person}{Peter
  Papcun}, \bibinfo{person}{Erik Kaj{\'a}ti}, \bibinfo{person}{Martin
  Mi{\v{s}}kuf}, {and} \bibinfo{person}{Jozef Mocnej}.}
  \bibinfo{year}{2018}\natexlab{}.
\newblock \showarticletitle{Smart and cognitive solutions for Operator 4.0:
  Laboratory H-CPPS case studies}.
\newblock \bibinfo{journal}{\emph{Computers {\&} Industrial Engineering}}
  (\bibinfo{year}{2018}), \bibinfo{pages}{105471}.
\newblock
\showISSN{0360-8352}
\urldef\tempurl%
\url{https://doi.org/10.1016/j.cie.2018.10.032}
\showDOI{\tempurl}


\end{thebibliography}

\end{document}